\documentclass[twocolumn]{svjour3}
\usepackage{amsmath,amssymb}
\usepackage{graphicx}
\usepackage{enumerate}
\usepackage{natbib}

\begin{document}

\title{Sampling from manifold-restricted distributions\\using tangent bundle projections}

\author{Alvin J. K. Chua}

\institute{Alvin J. K. Chua\\
\email{alvin.j.chua@jpl.nasa.gov} \at Jet Propulsion Laboratory, California Institute of Technology, Pasadena, CA 91109, U.S.A.}

\date{Received: date / Accepted: date}

\maketitle

\sloppy
\abovedisplayskip 6pt minus 2pt
\belowdisplayskip 6pt minus 2pt

\begin{abstract}
A common problem in Bayesian inference is the sampling of target probability distributions at sufficient resolution and accuracy to estimate the probability density, and to compute credible regions. Often by construction, many target distributions can be expressed as some higher-dimensional closed-form distribution with parametrically constrained variables, i.e., one that is restricted to a smooth submanifold of Euclidean space. I propose a derivative-based importance sampling framework for such distributions. A base set of $n$ samples from the target distribution is used to map out the tangent bundle of the manifold, and to seed $nm$ additional points that are projected onto the tangent bundle and weighted appropriately. The method essentially acts as an upsampling complement to any standard algorithm. It is designed for the efficient production of approximate high-resolution histograms from manifold-restricted Gaussian distributions, and can provide large computational savings when sampling directly from the target distribution is expensive.
\end{abstract}

\section{Introduction}
\label{sec:introduction}

Bayesian inference is a standard approach to the analysis of data in the physical sciences. The process involves fitting a probabilistic physical model to data, and summarizing the result as a probability distribution on the parameters of the model \citep{gelman2013bayesian}. In continuous problems, it entails the mapping out of a probability density function over the model parameter space. If the target distribution is nontrivial and the space has modest dimensionality, this must typically be performed with stochastic sampling algorithms, the most ubiquitous of which are the well-known Markov chain Monte Carlo (MCMC) class of methods \citep{robert2005monte,gamerman2006markov}. A histogram of the samples may then be used to estimate the probability density of the target distribution, and to compute credible regions in parameter space. However, the histogram bins must be small enough to resolve key features of the target density, which in turn calls for a large number of samples to ensure each bin gives an accurate estimate of the local density. Such a requirement can be computationally intensive, if not prohibitive.

I attempt to address this problem for any target distribution that can be cast as a higher-dimensional multivariate distribution with parametrized constraints on its variables. More precisely, the unnormalized target density must be expressible as the composition of (i) a smooth map from parameter space to an ambient Euclidean space, with (ii) the closed-form density of a common (e.g., multivariate normal) distribution on the ambient space. Many Bayesian likelihood functions ``factor'' naturally into such a composite form, with the map being a deterministic model for some set of observables, and the density describing the statistical uncertainty in their observed values. The present work includes some results for a general ambient density, but mostly focuses on the case where it is Gaussian. Even with this assumption, the smooth map is generally nonlinear, which endows its image with nontrivial curvature, and the full target density with non-Gaussianity.

In this paper, I propose a method that is designed for efficient sampling from distributions of the above form. It is original to the best of my knowledge, and combines a standard sampling algorithm such as Metropolis--Hastings (MH) \citep{doi:10.1093/biomet/57.1.97} with the derivative(s) of the smooth map, in order to perform importance sampling \citep{ripley1987stochastic} on a discretized and linearized approximation of the map image in the ambient space. The standard algorithm is used to explore the image space, which is a smooth submanifold of Euclidean space, and to populate a ``base chain'' of $n$ samples from the target distribution. Each base sample then seeds a ``mini-distribution'' of $m$ additional samples, where $m$ is specified; this is done by drawing ambient points from a suitably compact Gaussian centered on the mapped base sample, projecting them onto the associated tangent space, and assigning approximate weights to the pullback of these points onto parameter space. The result is a set of $nm$ weighted samples, from just $n$ calls to the map and derivative(s).

Provided first-derivative information is available, the proposed method serves as a runtime or post hoc ``multiplier'' for any algorithm that is employed to generate samples from a manifold-restricted Gaussian distribution. It is particularly useful when (i) the map is computationally expensive, (ii) the first derivative with respect to the $s$ parameters is not significantly more expensive than the map itself, and (iii) the base chain is near convergence. In this case, the computational savings scale as $\mathcal{O}(m/s)$ (where $m$ is effectively arbitrary), and the method can produce a high-resolution set of samples with great efficiency. Although these samples are approximate, the loss of accuracy is mitigated by the weights, and controlled by tuning the overall ``compactness'' of the mini-distributions; it can also be further reduced with the second derivative of the map, by accounting for the local curvature of the manifold in the compactness of each mini-distribution.

While the proposed method involves derivative-based sampling, it is really more of a bridge to kernel density estimation and other non-parametric approaches \citep{hjort2010bayesian}, and is conceptually distinct from MCMC samplers that use derivatives to inform the chain dynamics. There is a whole host of said samplers in the literature; prominent examples include Hamiltonian Monte Carlo (HMC) \citep{DUANE1987216}, the Metropolis-adjusted Langevin algorithm (MALA) \citep{doi:10.1111/1467-9868.00123}, and a family of stochastic-gradient MCMC techniques \citep{Ma:2015:CRS:2969442.2969566}. These typically rely only on the parameter derivatives of the scalar-valued probability density, rather than those of the vector-valued smooth map (from which the former may be derived). The method can still increase the computational efficiency of such algorithms due to its multiplicative effect, but no more than it would for a derivative-free sampler.

There are manifold-based MCMC samplers that do employ the full map derivatives in the construction of the chain; the proposed method is naturally paired with such algorithms, due to their shared computation of derivatives at each iteration. For example, Riemannian-manifold versions of HMC and MALA \citep{doi:10.1111/j.1467-9868.2010.00765.x,XIFARA201414} can make use of the pullback metric induced on parameter space, which is specified completely by the first derivative of the map. Manifold-constrained HMC methods \citep{pmlr-v22-brubaker12,2018arXiv180702356L} also work on the tangent bundle of the manifold, and hence require the first derivative to define a basis for each tangent space. More generally, any sampling of some probability distribution on a submanifold of Euclidean space can be informed by its density with respect to the Hausdorff measure on the submanifold, which may be computed using the first derivative as well \citep{diaconis}.

The proposed method bears some similarity to a Bayesian technique used in astronomy \citep{10.1046/j.1365-8711.2000.03692.x,10.1111/j.1365-2966.2005.09305.x}, where the gist is to sample from an ersatz likelihood that preserves the Fisher matrix at some chosen point, or from an unweighted ensemble of such likelihoods. This is essentially a simple projection of the target density onto an arbitrary set of tangent spaces, which can fail badly (as found by \cite{10.1111/j.1745-3933.2011.01034.x}). Finally, despite a common usage of techniques and nomenclature from differential geometry, the proposed method is only tangentially related to the premise and concepts of information geometry \citep{amari2007methods}, and also does not seem directly applicable to the computer science field of manifold learning \citep{lee2007nonlinear}---although it might be relevant for learning distributions on learned manifolds, e.g., as in \cite{arvanitidis2018latent}.

Section \ref{sec:method} details the theoretical and practical aspects of the method, which is then showcased through a handful of heuristic examples in Section \ref{sec:examples}; these range from the abstract to the applied (with a notable example in the presently high-profile field of gravitational-wave astronomy), while spanning various scenarios with different dimensionality setups and curvature profiles.

\section{Method}
\label{sec:method}

In Sections \ref{subsec:formalism} and \ref{subsec:gaussian}, I lay out some geometric preliminaries and present a formal derivation of the method. Readers who are more interested in practical implementation might benefit from the algorithmic summary provided in Section \ref{subsec:summary}. To the best of my ability, notation in this paper is chosen for intuitiveness, concordance with standard conventions, and prevention of symbolic overload. The Latin indices $i,j$ are used as set labels, while the Greek indices $\mu,\nu$ are reserved for local coordinates and the corresponding components of tensors.

\subsection{General formalism}
\label{subsec:formalism}

Let us consider the problem of generating random samples $\theta\in\mathrm{\Theta}$ from a specified probability distribution on the sample space $\mathrm{\Theta}$. We require that the target probability density is of the general form
\begin{equation}\label{eq:pdf}
p(\theta)\propto f(\alpha(\theta))
\end{equation}
with respect to the Lebesgue measure on $\mathrm{\Theta}$, where the map $\alpha:\mathrm{\Theta}\to\mathbb{R}^d$ is a smooth embedding of the sample-space manifold $\mathrm{\Theta}\cong\mathbb{R}^s$ (with $s<d$), and the ambient ``density'' $f:\mathbb{R}^d\to\mathbb{R}$ is a non-negative Lebesgue-integrable function. It is often more practical to take a bounded subset $\mathrm{\Theta}_b\subset\mathrm{\Theta}$ as the sample space instead, in which case the target density is of the form
\begin{equation}\label{eq:boundedpdf}
p(\theta)\propto f(\alpha(\theta))\mathbf{1}_b(\theta),
\end{equation}
where $\mathbf{1}_b$ is the indicator function of $\mathrm{\Theta}_b$.

The image manifold $\mathcal{M}:=\alpha[\mathrm{\Theta}]$ (or $\alpha[\mathrm{\Theta}_b]$) is then an $s$-dimensional submanifold of $\mathbb{R}^d$, and the pullback of the Euclidean metric by $\alpha$ induces a Riemannian metric $\mathrm{I}$ on $\mathrm{\Theta}$.\footnote{The notation $\mathrm{I}$ is historically tied to the first fundamental form (i.e., the special case of a surface in $\mathbb{R}^3$), but is chosen here for aesthetic consistency.} In local coordinates $\theta^\mu$, the components of this pullback metric are given by
\begin{equation}\label{eq:metric}
F_\mathrm{I}:=[\mathrm{I}(\partial_\mu,\partial_{\mu'})]=J^TJ,
\end{equation}
where $J$ is the $d\times s$ Jacobian matrix $[\partial_\mu\alpha^\nu]$. There is a natural inclusion of the tangent bundle $T\mathcal{M}$ into $\mathbb{R}^d\times\mathbb{R}^d$, and the columns of $J(\theta)$ form a basis in $\mathbb{R}^d$ for the corresponding tangent space $T_{\alpha(\theta)}\mathcal{M}\cong\mathbb{R}^s$.

In the proposed method, we assume that the target density can be sampled through other means, and that there exists a ``base chain'' with $n$ members:
\begin{equation}\label{eq:basechain}
B:=\{(\theta_i,\alpha_i,J_i)\,|\,i=1,2,\ldots,n\},
\end{equation}
where the set $\{\theta_i\}$ is distributed according to $p$, and $(\alpha_i,J_i):=(\alpha(\theta_i),J(\theta_i))$. For each triple in $B$, let us generate a ``mini-distribution'' of $m$ points $\beta_{ij}\in\mathbb{R}^d$, where $j=1,2,\ldots,m$. These are normally distributed with mean $\alpha_i$ and a covariance that depends on $\theta_i$, i.e.,
\begin{equation}\label{eq:gaussiansamples}
\beta_{ij}\sim\mathcal{N}\!\left(\alpha_i,C_i^{-1}\right),
\end{equation}
where $\beta$ is used here (and henceforth) to denote points that are not necessarily restricted to the manifold $\mathcal{M}$. The matrix $C_i$ determines the ``compactness'' of the points $\beta_{ij}$, and is left unspecified for now; a suitable prescription for choosing it is suggested in Section \ref{subsec:gaussian}.

If the second derivative of $\alpha$ is available, a natural choice for the compactness matrix $C$ will also account for the normal curvature of $\mathcal{M}$ at each point $\alpha_i$, i.e., $C$ should depend on the (generalized) second fundamental form $\mathrm{I\!I}$. This is a vector-valued form in the case of codimension $d-s>1$, and is given by the projection of the Euclidean directional derivative onto the normal space $N_{\alpha_i}\mathcal{M}$ \citep{spivak1970comprehensive}. We will find it convenient to treat $\mathrm{I\!I}$ as a vector in $\mathbb{R}^d$ rather than $N_{\alpha_i}\mathcal{M}$, and to work with the Euclidean norm $\left|\cdot\right|$ of $\mathrm{I\!I}$.\footnote{Such a representation is valid for general codimension, and obviates the need to construct a canonical basis for the normal bundle. Although the normal curvature is described more accurately by a signed norm with respect to the orientation of this basis, or even better by the full Riemann tensor, our approach is practically motivated and will typically be a conservative choice (see discussion around \eqref{eq:maxeigenvalue}).} This defines an $s\times s$ symmetric matrix
\begin{equation}\label{eq:ii}
F_\mathrm{I\!I}:=[|\mathrm{I\!I}(\partial_\mu,\partial_{\mu'})|]=\left[\left|\left(I-P^\perp\right)\nabla_{\partial_\mu}\partial_{\mu'}\right|\right],
\end{equation}
where $I$ is the identity matrix, $P^\perp(\theta_i)$ is the orthogonal projection onto $T_{\alpha_i}\mathcal{M}$, and $\nabla_vw$ denotes the directional derivative of $w$ along $v$ at $\alpha_i$ for $v,w\in T_{\alpha_i}\mathcal{M}$.

Eq. \eqref{eq:ii} can be cast explicitly in terms of $\partial\alpha$ and $\partial^2\alpha$, since the directional derivatives of the basis vectors are just the third-order Hessian tensor components
\begin{equation}\label{eq:hessian}
H^\nu_{\;\mu\mu'}:=\partial_\mu\partial_{\mu'}\alpha^\nu=\left(\nabla_{\partial_\mu}\partial_{\mu'}\right)^\nu.
\end{equation}
Additionally, the projection matrix is related to the Jacobian by \citep{meyer2000matrix}
\begin{equation}\label{eq:projectionoperator}
P^\perp=JJ^+,
\end{equation}
where the left pseudoinverse
\begin{equation}\label{eq:pseudoinverse}
J^+:=F_\mathrm{I}^{-1}J^T
\end{equation}
of $J$ also acts as the projection--pullback onto $\mathrm{\Theta}$.

We now use the well-known fact that for multivariate Gaussian random variables (on Euclidean space)
\begin{equation}\label{eq:generalgaussian}
(x^\mu,y^\nu)\sim\mathcal{N}(\mathbf{E}[(x^\mu,y^\nu)],\mathrm{Cov}(x^\mu,y^\nu)),
\end{equation}
the marginal distribution for any proper subset of variables $x^\mu$ is also Gaussian with unchanged covariance:
\begin{equation}\label{eq:marginalgaussian}
x^\mu\sim\mathcal{N}(\mathbf{E}[x^\mu],\mathrm{Cov}(x^\mu)).
\end{equation}
Hence the marginalization over $y^\nu$ of the Gaussian distribution in \eqref{eq:generalgaussian} is precisely its restriction to the subspace $y^\nu=\mathbf{E}[y^\nu]$ (which is flat with the Euclidean metric), and this restricted Gaussian may be sampled by orthogonally projecting a set of samples from \eqref{eq:generalgaussian} onto the subspace. The equivalence between flat-space restriction and orthogonal projection allows us to trivially generate samples from \eqref{eq:gaussiansamples} that are confined to the tangent space $T_{\alpha_i}\mathcal{M}$; these will then be taken as proxies for points on $\mathcal{M}$ in the neighborhood of $\alpha_i$.

For each Gaussian-distributed point $\beta_{ij}$ centered on $\alpha_i$, the projection of $\beta_{ij}-\alpha_i$ onto $T_{\alpha_i}\mathcal{M}$ gives the point
\begin{equation}\label{eq:projection}
\beta_{ij}^\perp:=\alpha_i+P_i^\perp[\beta_{ij}-\alpha_i],
\end{equation}
where $P_i^\perp:=P^\perp(\theta_i)$. Via pullback by $\alpha$, the associated point $\theta_{ij}\in\mathrm{\Theta}$ is given by
\begin{equation}\label{eq:projectionpullback}
\theta_{ij}=\theta_i+J_i^+[\beta_{ij}-\alpha_i],
\end{equation}
with $J_i^+:=J^+(\theta_i)$. In the neighborhood of $\alpha_i$, we have
\begin{equation}\label{eq:pushforward}
\alpha_{ij}:=\alpha(\theta_{ij})\approx\alpha_i+J_i[\theta_{ij}-\theta_i]=:\beta_i^\perp(\theta_{ij})=\beta_{ij}^\perp,
\end{equation}
with equality in the case where $\alpha$ is an affine map, i.e., when $\partial^2\alpha(\theta)=0$ for all $\theta\in\mathrm{\Theta}$.\footnote{Equivalently, when $\mathcal{M}$ is flat with the Euclidean metric.}

\begin{figure}[!tbp]
\centering
\includegraphics[width=\columnwidth]{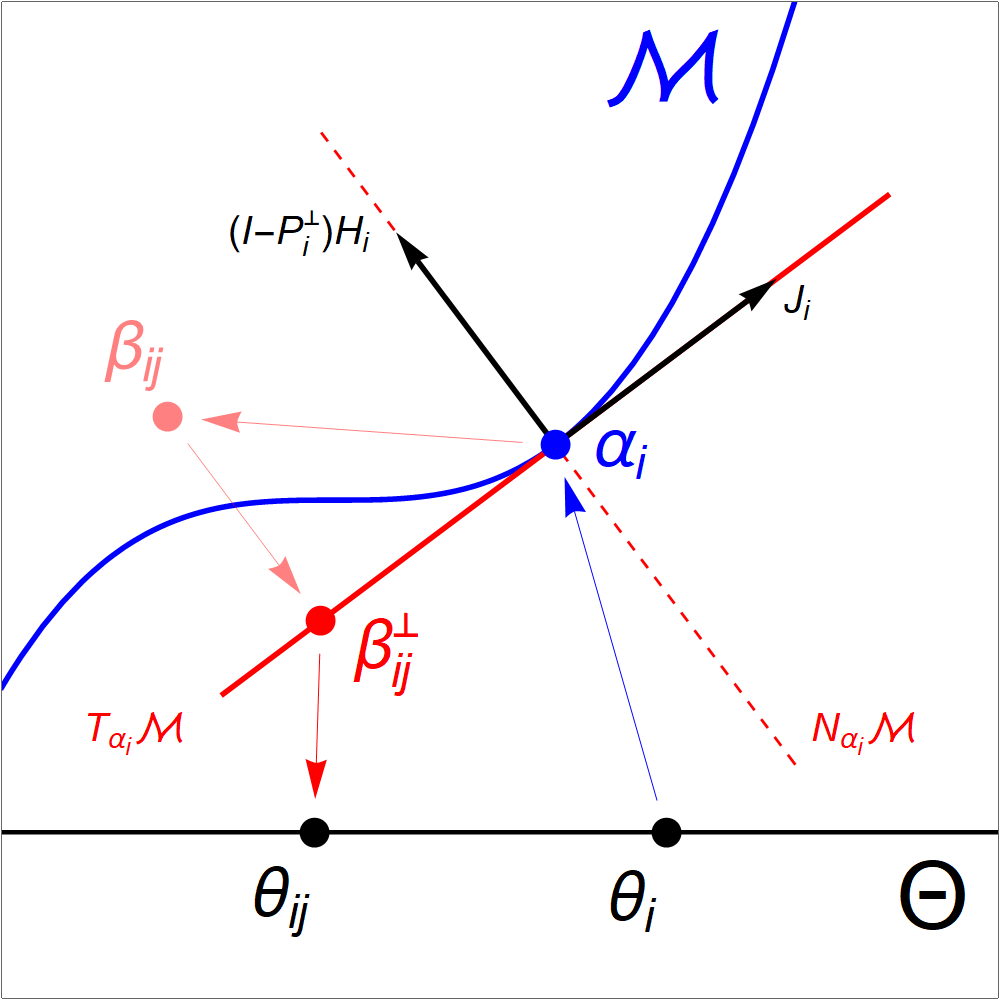}
\caption{Schematic depiction of various geometric concepts in Section \ref{subsec:formalism}. The sample $\theta_i\in\mathrm{\Theta}$ is mapped to the point $\alpha_i\in\mathcal{M}$, which seeds a random point $\beta_{ij}$ in the ambient space. Projecting $\beta_{ij}$ onto the tangent space at $\alpha_i$ gives the point $\beta_{ij}^\perp\in T_{\alpha_i}\mathcal{M}$, as well as the approximate sample $\theta_{ij}\in\mathrm{\Theta}$ via pullback. For a plane curve, $F_\mathrm{I}$ is the squared line element $|J_i|^2$, while $F_\mathrm{II}$ is the (unnormalized) curvature $|(I-P_i^\perp)H_i|$.}
\label{fig:dg}
\end{figure}

Figure \ref{fig:dg} depicts (for $(d,s)=(2,1)$) the various geometric concepts used to obtain a point on the tangent bundle of the manifold, along with the associated point in the sample space. However, as $\theta_{ij}$ is effectively drawn from a local mini-distribution conditioned on $\theta_i$, it must be assigned some appropriately defined importance weight $w(\theta_{ij})$ before it can be treated as an approximate sample from the target distribution. More precisely, we will need to estimate the density $q$ of the generating distribution for the set $\{\theta_{ij}\}$, and to choose
\begin{equation}\label{eq:optimalweights}
w_{ij}:=w(\theta_{ij})\approx\frac{p(\theta_{ij})}{q(\theta_{ij})}
\end{equation}
such that the expectation of a test function $\tau:\mathrm{\Theta}\to\mathbb{R}$ can be computed using $\{\theta_{ij}\}$ (in lieu of $\{\theta_i\}$):
\begin{align}\label{eq:testexpectation}
\mathbf{E}_p[\tau]&=\int_\mathrm{\Theta}d\theta\,\tau(\theta)p(\theta)\nonumber\\
&\approx\int_\mathrm{\Theta}d\theta\,\tau(\theta)w(\theta)q(\theta)=\mathbf{E}_q[\tau w].
\end{align}

From \eqref{eq:gaussiansamples} and \eqref{eq:projectionpullback}, we see that the conditional distribution of $\theta_{ij}$ given $\theta_i$ has the Gaussian density\footnote{If $x\sim\mathcal{N}(v,M)$, then $Ax+b\sim\mathcal{N}(Av+b,AMA^T)$.}
\begin{equation}\label{eq:conditionaldensity}
q'(\theta_{ij}|\theta_i)\propto\exp{\left(-\frac{1}{2}[\theta_{ij}-\theta_i]^TC'_i[\theta_{ij}-\theta_i]\right)},
\end{equation}
with $C'_i$ given by
\begin{equation}\label{eq:compactnesspullback}
C'_i=\left(J_i^+C_i^{-1}\left(J_i^+\right)^T\right)^{-1},
\end{equation}
i.e., $(C')^{-1}$ is the covariance from \eqref{eq:gaussiansamples} under local projection and pullback. The density for the marginal distribution of $\theta_{ij}$ is then given by
\begin{equation}\label{eq:marginaldensity}
q(\theta_{ij})=\int_\mathrm{\Theta}d\theta'\,q'(\theta_{ij}|\theta')p(\theta'),
\end{equation}
which is essentially the convolution of $p$ and $q'$ on $\mathrm{\Theta}$. This integral is in general both analytically and numerically intractable, due to its dependence on the arbitrary functions $\alpha(\theta)$, $f(\beta)$ and $C(\theta)$.

Regardless, let us now incorporate the linear approximation introduced in \eqref{eq:pushforward}, i.e., the use of pushforward points $\beta_i^\perp(\theta)\in T\mathcal{M}$ in place of mapped points $\alpha(\theta)\in\mathcal{M}$. Linearizing the optimal weights in \eqref{eq:optimalweights} about some fixed $\theta_i$, we set
\begin{equation}\label{eq:weights}
w_{ij}=\frac{\bar{p}(\theta_{ij})}{\bar{q}(\theta_{ij})},
\end{equation}
where the linearized densities $\bar{p}$ and $\bar{q}$ are given by
\begin{equation}\label{eq:linearp}
\bar{p}(\theta_{ij})\propto f\left(\beta^\perp(\theta_{ij})\right),
\end{equation}
\begin{equation}\label{eq:linearq}
\bar{q}(\theta_{ij})\propto\int_\mathrm{\Theta}d\theta'\,q'(\theta_{ij}|\theta')f\left(\beta^\perp(\theta')\right).
\end{equation}
With an appropriate choice of $C(\theta')$ in \eqref{eq:linearq}, the weights \eqref{eq:weights} do not depend on the map $\alpha$ beyond the point $\theta_i$ (as $J(\theta')=J_i$). This is the essence of the proposed method, since only $n$ calls to $\alpha$ and its derivative(s) are employed to generate a set of $nm$ weighted samples:
\begin{equation}\label{eq:samples}
S:=\{(\theta_{ij},w_{ij})\,|\,i=1,2,\ldots,n,\,j=1,\ldots,m\}.
\end{equation}

By construction, the weights \eqref{eq:weights} tend to the optimal weights \eqref{eq:optimalweights} in two separate limits, such that the approximation in \eqref{eq:testexpectation} becomes exact. The first is the flat-space limit, where $\alpha$ is an affine map; then $\beta_i^\perp=\alpha$ for all $i$, and we clearly have $\bar{p}/\bar{q}=p/q$ as required. Secondly, let us consider the smallest eigenvalue $c$ of $C$, and the case $c\to\infty$. In this limit of high compactness, we have $q'(\theta_{ij}|\theta_i)=\delta(\theta_{ij}-\theta_i)$ for all $i$, where $\delta$ is the Dirac delta function. Evaluating the integrals in \eqref{eq:marginaldensity} and \eqref{eq:linearq}, we find $q=p$ and $\bar{q}=\bar{p}$, such that $\bar{p}/\bar{q}=p/q$ again as required. In practical terms, these two limits imply that the systematic error in the method can be reduced by restricting to a more localized sample space (which will be closer to flat space), or by increasing the compactness of the geometrically generated samples (although they might then be not significantly more informative than the base samples).

Finally, we may quantify said error as the expected change in the expectation of the test function $\tau$, i.e., the error $\mathrm{\Delta}\mathbf{E}[\tau]$ incurred by using $\mathbf{E}_q[\tau w]$ to estimate $\mathbf{E}_p[\tau]$ in \eqref{eq:testexpectation}. Making the finite-difference approximations $p\approx\bar{p}(1+\mathrm{\Delta}\ln{\bar{p}})$ and $q\approx\bar{q}(1+\mathrm{\Delta}\ln{\bar{q}})$, we have
\begin{align}\label{eq:testerror}
\mathrm{\Delta}\mathbf{E}[\tau]&=\mathbf{E}_q[\tau w]-\mathbf{E}_p[\tau]\nonumber\\
&=\int_\mathrm{\Theta}d\theta\,\tau\left(\frac{\bar{p}}{\bar{q}}-\frac{p}{q}\right)q\nonumber\\
&\approx\int_\mathrm{\Theta}d\theta\,\tau\left(\frac{\mathrm{\Delta}\ln{\bar{q}}-\mathrm{\Delta}\ln{\bar{p}}}{1+\mathrm{\Delta}\ln{\bar{q}}}w\right)q\nonumber\\
&=:\mathbf{E}_q[\tau\mathrm{\Delta}w],
\end{align}
where $\theta$-dependence has been suppressed in notation, and the ``error weights'' $\mathrm{\Delta}w(\theta)$ are implicitly defined by the last equality. With the generality of the present setup, $\mathrm{\Delta}w$ is again intractable and \eqref{eq:testerror} only provides a theoretical estimate of the error. However, certain assumptions on the functional forms of $f$ and $C$ can admit an analytical approximation to $\mathrm{\Delta}w$, which might then allow the computation of $\mathrm{\Delta}\mathbf{E}[\tau]$ from the samples \eqref{eq:samples}.

\subsection{Manifold-restricted Gaussian}
\label{subsec:gaussian}

The proposed method is applicable in principle to the manifold restriction of any continuous multivariate probability distribution on Euclidean space (see \cite{kotz2004continuous} for examples). In practice, the convolution of the target density \eqref{eq:pdf} or \eqref{eq:boundedpdf} with the Gaussian conditional density \eqref{eq:conditionaldensity} must be derived or approximated analytically. Strategies for doing so might include computing the characteristic functions corresponding to $p$ and $q'$ \citep{shephard_1991} or, if the ambient density $f$ is near-Gaussian, adopting the formal approach used for path integrals in quantum field theory \citep{peskin1995introduction}. However, further treatment of $f$ in full generality is beyond the scope of the present work. 

For the rest of this paper, let us consider only the case where $f$ is proportional to the probability density of a Gaussian distribution $\mathcal{N}(\beta_*,\mathrm{\Sigma})$ on $\mathbb{R}^d$, i.e., we set
\begin{equation}\label{eq:gaussiandensity}
f(\beta)=\exp{\left(-\frac{1}{2}[\beta-\beta_*]^T\mathrm{\Sigma}^{-1}[\beta-\beta_*]\right)}.
\end{equation}
This is not an overly limiting assumption, since manifold-restricted Gaussians appear in a broad range of practical applications. For example, $f(\alpha(\theta))$ might be a Bayesian likelihood function where $\beta_*$ is some modeled observable $\alpha(\theta_*)$ with additive Gaussian noise $\nu$; it can then be sampled to map out the likelihood hypersurface over $\mathrm{\Theta}$, and to find the maximum-likelihood estimate $\theta_{**}\approx\theta_*+J^+_*\nu$. Usage of the method for Bayesian inference is elaborated on in Section \ref{subsec:summary}. In Section \ref{subsec:beta}, we will also look at reparametrization as a possible avenue towards generalizing the method for exponential families of distributions \citep{brown1986fundamentals}.

As is the case with Gaussians, the density \eqref{eq:pdf} with \eqref{eq:gaussiandensity} has several desirable properties. For one, the derivatives of $p(\theta)$ itself are quite tractable; in particular, the stationary points $\theta_s$ of $p$ occur where the vector $\mathrm{\Sigma}^{-1}[\alpha_s-\beta_*]$ is orthogonal to $T_{\alpha_s}\mathcal{M}$.\footnote{From $\partial p=f'(\alpha)\partial\alpha=0$, we see that $J^T_s\mathrm{\Sigma}^{-1}[\alpha_s-\beta_*]=0$.} The Fisher information matrix for $p$ also takes on the simple form
\begin{equation}\label{eq:fisher}
\mathrm{\Gamma}:=[\mathbf{E}[(\partial_\mu\ln{p})(\partial_{\mu'}\ln{p})]]=J^T\mathrm{\Sigma}^{-1}J,
\end{equation}
while the linearized target density \eqref{eq:linearp} expands to
\begin{equation}\label{eq:gaussianlinearp}
\bar{p}(\theta_{ij})\propto\exp{\left(-\frac{1}{2}\left[\beta_{ij}^\perp-\beta_*\right]^T\mathrm{\Sigma}^{-1}\left[\beta_{ij}^\perp-\beta_*\right]\right)}.
\end{equation}

Lastly, and most crucially, the convolution of the two Gaussians in the linearized generating density \eqref{eq:linearq} admits an analytic solution under two conditions: (i) if the sample space is $\mathrm{\Theta}$ and not $\mathrm{\Theta}_b$, and (ii) if the compactness $C$ is constant over $\mathrm{\Theta}$. Neither of these conditions is intended to be enforced in practice, and so the solution will only be an approximation (to the already approximate density $\bar{q}$). Regardless, we find
\begin{equation}\label{eq:gaussianlinearq}
\bar{q}(\theta_{ij})\propto\exp{\left(-\frac{1}{2}\left[\beta_{ij}^\perp-\beta_*\right]^T(\mathrm{\Sigma}'_i)^{-1}\left[\beta_{ij}^\perp-\beta_*\right]\right)},
\end{equation}
with $\mathrm{\Sigma}'_i$ given by \citep{bishop2013pattern}
\begin{equation}\label{eq:Sigma_i}
\mathrm{\Sigma}'_i=\mathrm{\Sigma}+J_i (C'_i)^{-1}J_i^T=\mathrm{\Sigma}+P_i^\perp C_i^{-1}P_i^\perp.
\end{equation}
Eq. \eqref{eq:gaussianlinearq} can be almost exact even when restricting to $\mathrm{\Theta}_b$, provided the bulk of the density \eqref{eq:pdf} is contained within the boundary of $\mathrm{\Theta}_b$. Boundary corrections when this is not the case are discussed in Section \ref{subsec:summary}.

From \eqref{eq:gaussianlinearp} and \eqref{eq:gaussianlinearq}, the weights \eqref{eq:weights} simplify to
\begin{equation}\label{eq:gaussianweights}
w_{ij}=N_i\exp{\left(-\frac{1}{2}\left[\beta_{ij}^\perp-\beta_*\right]^TM_i\left[\beta_{ij}^\perp-\beta_*\right]\right)},
\end{equation}
with $M_i$ and $N_i$ given by
\begin{equation}\label{eq:M_i}
M_i=\mathrm{\Sigma}^{-1}-\left(\mathrm{\Sigma}+P_i^\perp C_i^{-1}P_i^\perp\right)^{-1},
\end{equation}
\begin{equation}\label{eq:N_i}
N_i=\sqrt{\frac{\det{\left(J_i^+\left(\mathrm{\Sigma}+C_i^{-1}\right)\left(J_i^+\right)^T\right)}}{\det{\left(J_i^+\mathrm{\Sigma}\left(J_i^+\right)^T\right)}}},
\end{equation}
where \eqref{eq:N_i} is estimated using $\mathrm{\Sigma}$ and $\mathrm{\Sigma}'$ under local projection and pullback (as $\bar{p}$ and $\bar{q}$ are defined on $\mathrm{\Theta}$ and not $\mathbb{R}^d$). Note that variability over $\mathrm{\Theta}$ has been restored to $C$ in \eqref{eq:gaussianlinearq}--\eqref{eq:N_i}, through its implicit dependence on $\theta_i$. This choice is empirically motivated, since fixing the compactness for a manifold with varying curvature generally incurs a larger penalty to the overall accuracy than adopting a location-dependent compactness.

Let us now construct a compactness matrix $C_i$ that depends on the minimal length scales describing (i) the ambient density $f$, (ii) the normal curvature of the manifold $\mathcal{M}$, and (iii) the extent of some sampling region in $\mathrm{\Theta}$ (i.e., the sample space $\mathrm{\Theta}_b$). The square of the first is simply the smallest eigenvalue $\sigma^2$ of the ambient covariance $\mathrm{\Sigma}$, while the reciprocal of the second is the largest eigenvalue $\kappa$ of a curvature-related $s\times s$ matrix
\begin{equation}\label{eq:curvature}
K:=Q^TF_\mathrm{I\!I}Q.
\end{equation}
Here $Q$ defines a change of basis such that $F_\mathrm{I}\to I$, and is given via the standard eigenvalue decomposition
\begin{equation}\label{eq:eigendecomposition}
F_\mathrm{I}^{-1}=UDU^T,
\end{equation}
\begin{equation}\label{eq:changeofbasis}
Q=UD^{1/2},
\end{equation}
with orthogonal $U$ and diagonal, positive-definite $D$.

The eigenvalues of $K$ are the unsigned norms of $\mathrm{I\!I}$ with respect to some tangent space eigenbasis, and reduce by design to more familiar expressions of curvature in various special cases. When $s=1$, $\mathcal{M}$ is a curve $\gamma(t)$ in Euclidean space, and the single eigenvalue is its curvature $\kappa(t)=|\gamma''(t)|$, with $t$ chosen such that $|\gamma'(t)|=1$ \citep{struik1961lectures}. In the case of a surface in $\mathbb{R}^3$ with positive curvature (both eigenvalues have the same sign), $K$ has the same determinant and eigenvalues as the Weingarten map, i.e., the Gaussian and principal curvatures respectively \citep{struik1961lectures}.

More generally, $\kappa\geq0$ is proportional to an upper bound for the maximal eigenvalue $\kappa'\geq0$ of \eqref{eq:ii} with a signed norm (assuming the condition number of $F_\mathrm{I}$ varies slowly across the manifold). This is because
\begin{equation}\label{eq:maxeigenvalue}
\kappa'/\iota_\mathrm{max}^2\leq u^TF_\mathrm{I\!I}'u\leq v^TF_\mathrm{I\!I}v\leq w^TF_\mathrm{I\!I}w\leq\kappa/\iota_\mathrm{min}^2,
\end{equation}
where $\iota_\mathrm{min}/\iota_\mathrm{max}$ is the condition number of $F_\mathrm{I}$, the matrix $F_\mathrm{I\!I}'$ is given by \eqref{eq:ii} but evaluated using a signed norm, the unit vectors $u,w$ are eigenvectors of $F_\mathrm{I\!I}',F_\mathrm{I\!I}$ respectively, and the unit vector $v$ has components $v^\mu=|u^\mu|$. In other words, an appropriately scaled $\kappa$ is conservative in regions of negative curvature, as it might overestimate the maximal principal curvature when the local geometry of the manifold is hyperbolic.

For simplicity, we will assume a hyperrectangular sampling region $\mathrm{\Theta}_b$ with sides of half-length $\ell_1,\ldots,\ell_s$. These determine the associated matrix
\begin{equation}\label{eq:boxlengths}
L:=\mathrm{diag}\left(\ell_1^{-2},\ldots,\ell_s^{-2}\right),
\end{equation}
its pushforward by $\alpha$:
\begin{equation}\label{eq:boxpushforward}
\mathrm{\Lambda}:=(J^+)^TLJ^+,
\end{equation}
and the largest eigenvalue $\lambda^2$ of $\mathrm{\Lambda}$. A practical choice for the compactness is then $C_i=c_iI$, with $c_i$ given by
\begin{equation}\label{eq:compactness}
c_i=\frac{1}{\epsilon}\left(\max{\left\{\lambda_i^2,\frac{\kappa_i}{\sigma}\right\}}\right),
\end{equation}
where the dimensionless sampling parameter $\epsilon>0$ is tuned at runtime, and $(\lambda_i,\kappa_i):=(\lambda(\theta_i),\kappa(\theta_i))$.

The curvature term $\kappa_i/\sigma$ in \eqref{eq:compactness} is derived by constraining the displacement due to curvature that is associated with the length $1/\sqrt{c_i}$ in $T_{\alpha_i}\mathcal{M}$, i.e., by requiring that the largest distance between corresponding points on $T_{\alpha_i}\mathcal{M}$ and $\mathcal{M}$ is proportional to the smallest covariance length of the ambient density (which is constant). Consequently, it might be useful to think of $\epsilon\sigma$ as an approximate upper bound on the error $|\beta_{ij}^\perp-\alpha_{ij}|$, for $\beta_{ij}$ at one (mini-distribution) sigma. The metric-only term $\lambda_i^2$ in \eqref{eq:compactness} is typically less stringent than the curvature term, but might perform a similar function if the Hessian $H$ is not readily available, since the metric contributes to and can be strongly correlated with the normal curvature. This term primarily safeguards against any regions where $\kappa_i$ and the Fisher information $\det{(\mathrm{\Gamma})}$ both become small, and is crucial for the example in Section \ref{subsec:reparabola}, where $\mathcal{M}$ has vanishing normal curvature but the metric on $\mathrm{\Theta}$ has singular points. 

It is convenient to cast \eqref{eq:gaussiandensity} in ``whitened'' form via the linear transformation $\beta\to W\beta$ (where $W^TW=\mathrm{\Sigma}^{-1}$ for some preferred choice of $W$ \citep{doi:10.1080/00031305.2016.1277159}), such that $\mathrm{\Sigma}\to I$ and the length scales of the ambient density are encoded in the transformed embedding $W\alpha$. The scalar compactness \eqref{eq:compactness} reduces to
\begin{equation}\label{eq:simplecompactness}
c_i=\frac{1}{\epsilon}\left(\max{\left\{\lambda_i^2,\kappa_i\right\}}\right),
\end{equation}
while $M_i$ and $N_i$ in the Gaussian weights \eqref{eq:gaussianweights} become
\begin{equation}\label{eq:simpleM_i}
M_i=\frac{1}{1+c_i}P_i^\perp,
\end{equation}
\begin{equation}\label{eq:simpleN_i}
N_i=\left(\frac{1+c_i}{c_i}\right)^{s/2},
\end{equation}
where \eqref{eq:simpleM_i} follows from the matrix inversion lemma \citep{press2007numerical} and the decomposition \eqref{eq:projectionoperator}.

Note that with $C_i=c_iI$ and $\mathrm{\Sigma}=I$, the geometric generation of the points $\theta_{ij}$ is conceptually equivalent to the sampling of a Fisher-based mini-distribution:
\begin{equation}\label{eq:fishersamples}
\theta_{ij}\sim\mathcal{N}\left(\theta_i,\frac{1}{c_i}\mathrm{\Gamma}_i^{-1}\right),
\end{equation}
where the covariance has been scaled such that the pushforward points $\beta_i^\perp(\theta_{ij})$ lie at a controlled distance from the manifold. This projection-free prescription might be convenient in practice, e.g., if the base chain is being constructed concurrently from a similar (unscaled) MH proposal distribution; however, it brings no real computational benefits even when $d\gg s$, as the weights \eqref{eq:gaussianweights} must still be computed using $\beta_{ij}^\perp$.

We now return to the systematic error \eqref{eq:testerror} in the expectation of the test function, which may be evaluated further following the developments in Section \ref{subsec:gaussian}. Expanding the map $\alpha$ to second order about some fixed $\theta_i$, we write (using Einstein notation)
\begin{equation}\label{eq:quadalpha}
\tilde{\alpha}(\theta):=(\alpha_*)^\nu+J^\nu_{\;\mu}\vartheta^\mu+\frac{1}{2}H^\nu_{\;\mu\mu'}\vartheta^\mu\vartheta^{\mu'},
\end{equation}
\begin{equation}\label{eq:quadJ}
\tilde{J}(\theta):=J^\nu_{\;\mu}+H^\nu_{\;\mu\mu'}\vartheta^{\mu'},
\end{equation}
where $\alpha_*:=\alpha_i-\beta_*$ and $\vartheta:=\theta-\theta_i$. Under the assumption that the Gaussian form of the linearized convolution \eqref{eq:gaussianlinearq} holds for $q$ at second order, the covariance at equivalent order is approximately $(I-\tilde{M})^{-1}$, where
\begin{equation}\label{eq:quadM}
\tilde{M}=\frac{1}{1+c}\tilde{J}\mathrm{\Gamma}^{-1}\tilde{J}^T.
\end{equation}

With the tilded quantities \eqref{eq:quadalpha}--\eqref{eq:quadM} defining the Gaussian second-order densities $\tilde{p}$ and $\tilde{q}$, we may set $\mathrm{\Delta}\ln{\bar{p}}\approx\ln{\tilde{p}}-\ln{\bar{p}}$ and $\mathrm{\Delta}\ln{\bar{q}}\approx\ln{\tilde{q}}-\ln{\bar{q}}$ in \eqref{eq:testerror}. The error weights at equivalent order then evaluate to
\begin{equation}\label{eq:errorweights}
\mathrm{\Delta}w\approx\frac{\mathrm{\Delta}_M}{\mathrm{\Delta}_M+(1+c)(1-\mathrm{\Delta}_I)}w,
\end{equation}
with $\mathrm{\Delta}_M$ and $\mathrm{\Delta}_I$ given by
\begin{align}\label{eq:DeltaM}
\mathrm{\Delta}_M=\;&(\alpha_*)^\nu J_{\nu\mu}\left(\mathrm{\Gamma}^{-1}\right)^{\mu\mu'}(\alpha_*)^{\nu'}H_{\nu'\mu'\mu''}\vartheta^{\mu''}\nonumber\\
&+\frac{3}{2}(\alpha_*)^\nu J_{\nu\mu}\left(\mathrm{\Gamma}^{-1}\right)^{\mu\mu'}J^{\nu'}_{\;\mu'}H_{\nu'\mu''\mu'''}\vartheta^{\mu''}\vartheta^{\mu'''}\nonumber\\
&+\frac{1}{2}(\alpha_*)^\nu H_{\nu\mu\mu'}\vartheta^{\mu'}\left(\mathrm{\Gamma}^{-1}\right)^{\mu\mu'}(\alpha_*)^{\nu'}H_{\nu'\mu'\mu''}\vartheta^{\mu''}\nonumber\\
&+(\alpha_*)^\nu H_{\nu\mu\mu'}\vartheta^\mu\vartheta^{\mu'},
\end{align}
\begin{equation}\label{eq:DeltaI}
\mathrm{\Delta}_I=\frac{1}{2}(\alpha_*)^\nu H_{\nu\mu\mu'}\vartheta^\mu\vartheta^{\mu'}.
\end{equation}
While somewhat unwieldy, these error weights can be evaluated for the samples \eqref{eq:samples} to provide a ballpark estimate of the error \eqref{eq:testerror} (as demonstrated in Section \ref{subsec:parabola}). They also explicitly reflect the limiting behaviour discussed in Section \ref{subsec:formalism}; we see that $\mathrm{\Delta}w=0$ in flat space since both $\mathrm{\Delta}_M$ and $\mathrm{\Delta}_I$ vanish identically, while $\mathrm{\Delta}w\to0$ as $c\to\infty$ in the limit of high compactness.\footnote{This limit is strictly ill defined if $\mathrm{\Delta}_I\to1$ as $c\to\infty$, but note that we also have $\mathbf{E}[\mathrm{\Delta}_I]\to0$ as $c\to\infty$.}

\subsection{Summary and usage}
\label{subsec:summary}

Starting with a base chain of $n$ samples $\theta_i$ from some manifold-restricted Gaussian distribution, the proposed method upsamples this to a set of $nm$ weighted samples $(\theta_{ij},w_{ij})$ from approximately the same distribution, where $m$ is a desired multiplier for the sampling resolution. Any standard algorithm may be employed to draw the base samples according to the target density $p(\theta)\propto f(\alpha(\theta))$, and it will also provide the corresponding points $\alpha_i$ on the manifold as a by-product of computing $p$. The first derivatives $J_i=\partial\alpha_i$ (and the second derivatives $H_i=\partial^2\alpha_i$, if available) can be evaluated either concurrently with $(\theta_i,\alpha_i)$ or post hoc; in the former case, they might be used to propose new base points in a MH algorithm, e.g., by taking the inverse Fisher matrix as the proposal covariance, or by adding a deterministic drift term as in manifold MALA.

Let us assume that the Gaussian density \eqref{eq:gaussiandensity} is in whitened form such that $\mathrm{\Sigma}=I$, and that the base chain is obtained through a derivative-free sampling algorithm. For each pair $(\theta_i,\alpha_i)$ in the base chain, we:

\begin{enumerate}
\item Evaluate the Jacobian $J_i$ (and the Hessian $H_i$).
\item Compute the metric $(F_\mathrm{I})_i$ from Eq. \eqref{eq:metric}, and the pseudoinverse $J_i^+$ from Eq. \eqref{eq:pseudoinverse}.
\item Compute the metric term $\lambda_i^2$ in Eq. \eqref{eq:simplecompactness} from Eqs \eqref{eq:boxlengths} and \eqref{eq:boxpushforward}. (Also compute the second fundamental form $(F_\mathrm{I\!I})_i$ from Eqs \eqref{eq:ii} and \eqref{eq:hessian}, and the curvature term $\kappa_i$ in Eq. \eqref{eq:simplecompactness} from Eqs \eqref{eq:curvature}--\eqref{eq:changeofbasis}.)
\item Draw the mini-distribution samples $\beta_{ij}$ in Eq. \eqref{eq:gaussiansamples}.
\item For each $\beta_{ij}$, compute the corresponding projection--pullback sample $\theta_{ij}$ from Eq. \eqref{eq:projectionpullback}.
\item For each $\theta_{ij}$, compute the corresponding pushforward point $\beta_{ij}^\perp$ from Eq. \eqref{eq:pushforward}, and the corresponding weight $w_{ij}$ from Eqs \eqref{eq:gaussianweights}, \eqref{eq:simpleM_i} and \eqref{eq:simpleN_i}.
\end{enumerate}

The proposed method is efficient when the computational complexity of each iteration is dominated by the evaluations of the $d$-vector $\alpha_i$ and its derivative(s) with respect to the $s$ parameters. In such cases, Step 1 is the rate-determining step of each iteration; it is computed at $\mathcal{O}(s)$ times the cost of a single $\alpha$-evaluation (or $\mathcal{O}(s^2)$ if $H_i$ is used), where the scaling coefficient can be small, e.g., by leveraging parallel computing in the evaluation of $J_i$. The method then has similar complexity to the manifold-based algorithms mentioned in Section \ref{sec:introduction}, and is ideally used in conjunction with them. It incurs some additional cost when combined with more standard derivative-based samplers, since only the gradient of $\ln{p}$ is typically required for such algorithms, and this is computed at $\mathcal{O}(1)$ times the cost of a single $\alpha$-evaluation. However, given the multiplicative nature of the method, the effective computational savings in obtaining a set of $nm$ samples still scale as $\mathcal{O}(m/s)$.

The remaining steps all comprise straightforward mathematical operations, but scale with some powers of $s$, $d>s$ (possibly $d\gg s$) and $m$. Step 2 only contains two evaluations of complexity $\mathcal{O}(ds^2)$, with the projection matrix \eqref{eq:projectionoperator} not explicitly required by the algorithm. Although the metric term in Step 3 is overtly $\mathcal{O}(d^3)$ as it involves the eigenvalue decomposition of the $d\times d$ matrix $\mathrm{\Lambda}$, the nonzero eigenvalues of this low-rank matrix are the squared singular values of the matrix $L^{1/2}J_i^+$, which may be obtained at only $\mathcal{O}(ds^2)$ cost \citep{golub1965calculating}. The second-order part of Step 3 is dominated by the computation of the second fundamental form, which is an $\mathcal{O}(ds^3)$ operation. Step 4 has complexity $\mathcal{O}(md)$, while Steps 5 and 6 are both $\mathcal{O}(mds)$ (with the quadratic form in \eqref{eq:gaussianweights} reduced to a scalar product by the decomposition \eqref{eq:projectionoperator}).

It is worth emphasizing here that while the proposed method might improve the rate of sampling convergence locally (i.e., in the vicinity of an isolated probability mode), it is not designed to address global convergence; indeed, the assumption throughout Sections \ref{subsec:formalism} and \ref{subsec:gaussian} is that the distribution of the base chain has already converged to the target distribution. Hence the performance of the method is largely independent of $m$ if the base chain is near convergence, and $\epsilon$ becomes the sole tuning parameter. If $\epsilon$ is set small, the distribution of the weighted samples is guaranteed to converge as $n\to\infty$, but the sampling resolution is tied to that of the base chain itself. Conversely, large $\epsilon$ results in improved resolution at the potential cost of reduced sampling accuracy (due to the systematic error \eqref{eq:testerror}).

One possible strategy for tuning $\epsilon$ is to start with some conservative choice $\epsilon\lesssim\max{\{\lambda_i^2,\kappa_i\}}$; this can then be increased if the mini-distributions are too tightly clustered around their base points, or decreased if the error $|\beta_{ij}^\perp-\alpha_{ij}|$ for an examined sample $\theta_{ij}$ is over some threshold. Since $1/\epsilon$ is essentially an overall scale for the compactness, computing the curvature term $\kappa_i$ becomes less important for a nearly constant Hessian in the sampling region, i.e., when $\partial^3\alpha_i\approx0$ for all $i$. Furthermore, if the Jacobian also varies slowly across the region of interest, we may simply take $c_i=1/\epsilon$ and do away with Step 3 altogether.

An additional procedure is required if the target density is non-negligible at the boundary of the sampling region $\mathrm{\Theta}_b$. From \eqref{eq:boundedpdf}, \eqref{eq:linearp} and \eqref{eq:linearq}, the linearized weights for a bounded sample space can be written as
\begin{equation}\label{eq:boundedweights}
w_b(\theta_{ij})=\frac{\bar{p}(\theta_{ij})\mathbf{1}_b(\theta_{ij})}{\int_\mathrm{\Theta}d\theta'\,q'(\theta_{ij}|\theta')\bar{p}(\theta')\mathbf{1}_b(\theta')},
\end{equation}
where the linearized generating density (the denominator) is the analytically nontrivial convolution of a truncated multivariate Gaussian with a regular one. There are several ways to sidestep the computation of \eqref{eq:boundedweights}; one that involves minimal modification to the existing framework makes use of the observation that the original weights \eqref{eq:weights} remain a good approximation to \eqref{eq:boundedweights} (both are $\sim1$) when the length scales associated with $q'$ are much smaller than those of $\mathrm{\Theta}_b$---as is typically the case for a conservative choice of $\epsilon$.

This weight approximation alone does not account for the fact that base samples just inside the boundary might seed projected points that fall outside the sampling region. Simply discarding such points will give rise to a deficit of samples in the affected mini-distributions, which might skew the overall distribution significantly. More precisely, projected samples just inside the boundary will effectively be generated according to a density that is artificially diminished there,
such that their assigned weights become less adequate. The most direct solution is to replace any projected point $\theta_{ij}\notin\mathrm{\Theta}_b$ with a unit-weight copy of its associated base sample $\theta_i$, which is formally equivalent to generating it in the limit of high compactness (since $(\theta_{ij},w_{ij})\to(\theta_i,1)$ as $c_i\to\infty$). To see that this is valid on the same level as the proposed method, note that the set of samples with replacement is effectively that obtained from a base chain $\{\vartheta_{i'}\}$ of length $n'=nm$ with mini-distributions of size $m'=1$, where $\vartheta_{i'}=\theta_{(i'\,\mathrm{mod}\,n)+1}$ and $c_{i'}\to\infty$ for some $\vartheta_{i'}$ near the boundary (selected a posteriori).

The approach described so far is ad hoc and might potentially reduce sampling resolution near the boundary, but it does preserve the overall distribution to a good approximation, and is also straightforward to perform during runtime or in post-processing. For simplicity, such boundary corrections are applied to the relevant examples in Sections \ref{subsec:klein}, \ref{subsec:gw} and \ref{subsec:beta}. An alternative, more involved treatment is to extend the framework to a target density of the form $p\propto(f\circ\alpha)g$, where $g$ is the probability density of some multivariate normal distribution $\mathcal{N}(\hat{\theta},\hat{\sigma})$ on $\mathrm{\Theta}$. This can be cast in the original form of \eqref{eq:pdf} with \eqref{eq:gaussiandensity}, by defining an augmented map $\hat{\alpha}$ into the larger ambient space $\mathbb{R}^{d+s}$, i.e., $\hat{\alpha}(\theta):=\alpha(\theta)\oplus\theta$. With the analogous augmentations $\hat{\beta}_*:=\beta_*\oplus\hat{\theta}$ and $\hat{\mathrm{\Sigma}}:=\mathrm{Diag}(\mathrm{\Sigma},\hat{\sigma})$ (where $\mathrm{Diag}$ here denotes a block diagonal matrix), we have
\begin{equation}\label{eq:augmentedgaussian}
p(\theta)\propto\exp{\left(-\frac{1}{2}\left[\hat{\alpha}(\theta)-\hat{\beta}_*\right]^T\hat{\mathrm{\Sigma}}^{-1}\left[\hat{\alpha}(\theta)-\hat{\beta}_*\right]\right)},
\end{equation}
in concordance with \eqref{eq:pdf} and \eqref{eq:gaussiandensity}. This approach comes at the cost of a higher-dimensional embedding, although the additional partial derivatives are trivial to compute. It can then be applied specifically to the density \eqref{eq:boundedpdf} with a hyperrectangular $\mathrm{\Theta}_b$, by choosing $(\hat{\theta},\hat{\sigma})=(0,I)$ and making the transformations $\theta^\mu\to\mathrm{\Phi}(\theta^\mu)$, where $\mathrm{\Phi}$ is the standard normal cumulative distribution function (up to scaling and translation).

In the context of Bayesian inference, the above discussion relates to the sampling of a posterior distribution with density $p\propto(f\circ\alpha)\pi$, where $f\circ\alpha$ is now a likelihood function and $\pi$ is some prior density on $\mathrm{\Theta}$. We have only considered uniform ($\pi\propto\mathbf{1}_b$) and Gaussian ($\pi=g$) priors thus far, but these might not be suitable in more general applications. The method can be formally extended to a broader class of priors, via an initial transformation $\iota(\theta')$ that maps the uniform distribution on a unit hypercube $\mathrm{\Theta}_h$ to the desired prior on $\mathrm{\Theta}$. Then $\iota$ defines a transformed map $\alpha_h:=\alpha\circ\iota$, such that the original (uniform-prior) framework may still be used to generate samples from the target posterior distribution, according to the transformed density
\begin{equation}\label{eq:pdfprior}
p_h(\theta')\propto f(\alpha_h(\theta'))\mathbf{1}_h(\theta')\propto f(\alpha(\theta))\pi(\theta)
\end{equation}
with respect to the Lebesgue measure on $\mathrm{\Theta}_h$. The map $\iota$ is known as the inverse Rosenblatt transformation \citep{rosenblatt1952}; it depends on the quantile function of the desired prior, which is generally not available in closed form but can be evaluated numerically for many standard distributions \citep{press2007numerical}.

Finally, a straightforward redefinition of the importance weights could also provide a post hoc treatment of generic priors. For some target posterior density $p_\pi\propto(f\circ\alpha)\pi$, let us first generate both a base chain and a set of weighted samples according to $p\propto f\circ\alpha$, as per the original framework. The linearized importance weights for $p_\pi$ are then given by
\begin{align}\label{eq:posthocweights}
w_\pi(\theta_{ij})&=\frac{\bar{p}_\pi(\theta_{ij})}{\int_\mathrm{\Theta}d\theta'\,q'(\theta_{ij}|\theta')\bar{p}(\theta')}\nonumber\\&\approx\frac{\pi(\theta_{ij})w(\theta_{ij})}{\int_\mathrm{\Theta}d\theta'\,\pi(\theta')w(\theta')\bar{q}(\theta')},
\end{align}
where the linearized generating density (the denominator in the first line) does not depend on $\pi$ since the base chain is distributed according to $p$. These ``prior weights'' may be computed from the sample set and initial weights as $w_\pi(\theta_{ij})\approx\pi(\theta_{ij})w_{ij}/\sum_{ij}\pi(\theta_{ij})w_{ij}$, provided the prior is not so informative that the majority of samples lie beyond the bulk of $\pi$.

\section{Examples}
\label{sec:examples}

The efficacy of the proposed method is now demonstrated through a number of heuristic examples: a parabola (Section \ref{subsec:parabola}); a family of Klein bottles (Section \ref{subsec:klein}); a space of gravitational-wave signals (Section \ref{subsec:gw}); a reparametrized parabola (Section \ref{subsec:reparabola}); and a reparametrized beta distribution (Section \ref{subsec:beta}). In each example, benchmarking is performed by way of high-resolution histograms, which are constructed for the target distribution (or an accurate reference chain) and the weighted samples produced through the method.

As a standard metric for the concurrence of two histograms with identical binning, we will use the Hellinger distance $d_H$ between two discrete probability distributions $\{p_i\},\{q_i\}$. This is given by \citep{pollard2002user}
\begin{equation}\label{eq:hellinger}
d_H=\sqrt{\frac{1}{2}\sum_i\left(\sqrt{p_i}-\sqrt{q_i}\right)^2},
\end{equation}
with the maximum distance achieved when there is no bin $i$ such that $p_iq_i>0$. The Hellinger distance comes with ease of interpretation, as it is symmetric and satisfies $0\leq d_H\leq1$; it is preferred here to the more commonly used Kullback--Leibler divergence \citep{pollard2002user}, which is unsuitable for histograms with a large number of empty bins. That being said, any choice of statistical distance between histograms will naturally be dependent on binning, and hence less meaningful for comparison across problems.

\subsection{Parabola}
\label{subsec:parabola}

We begin with a simple but instructive example: a bivariate normal distribution restricted to a parabola in the plane, such that $(d,s)=(2,1)$. With Cartesian coordinates $(x,y)$ on $\mathbb{R}^2$, we set
\begin{equation}\label{eq:parabolaalpha}
\alpha(x)=\left(x,x^2\right),
\end{equation}
\begin{equation}\label{eq:parabolabeta}
\beta_*=(1,2)
\end{equation}
in \eqref{eq:pdf} and \eqref{eq:gaussiandensity}. We also take $\mathrm{\Sigma}=I$ here (and in all of the other examples), such that the target density is
\begin{equation}\label{eq:parabolapdf}
p(x)\propto\exp{\left(-\frac{1}{2}\left(x^4-3x^2-2x+5\right)\right)}.
\end{equation}

The bulk of \eqref{eq:parabolapdf} lies comfortably within the interval $[-3,3]$, which we use as our sampling region; the metric and curvature terms in \eqref{eq:simplecompactness} then satisfy
\begin{equation}\label{eq:parabolacompactness}
\lambda_i^2=\frac{1}{9\left(1+4x_i^2\right)}<\frac{2}{\left(1+4x_i^2\right)^{3/2}}=\kappa_i
\end{equation}
for all $x_i\in[-3,3]$, such that $c_i=\kappa_i/\epsilon$. Note that as the second derivative of \eqref{eq:parabolaalpha} is constant, the $x$-dependence in the curvature is due to the changing metric, and so $\lambda_i^2$ is correlated with (and can be used in lieu of) $\kappa_i$.

\begin{figure}[!tbp]
\centering
\includegraphics[width=\columnwidth]{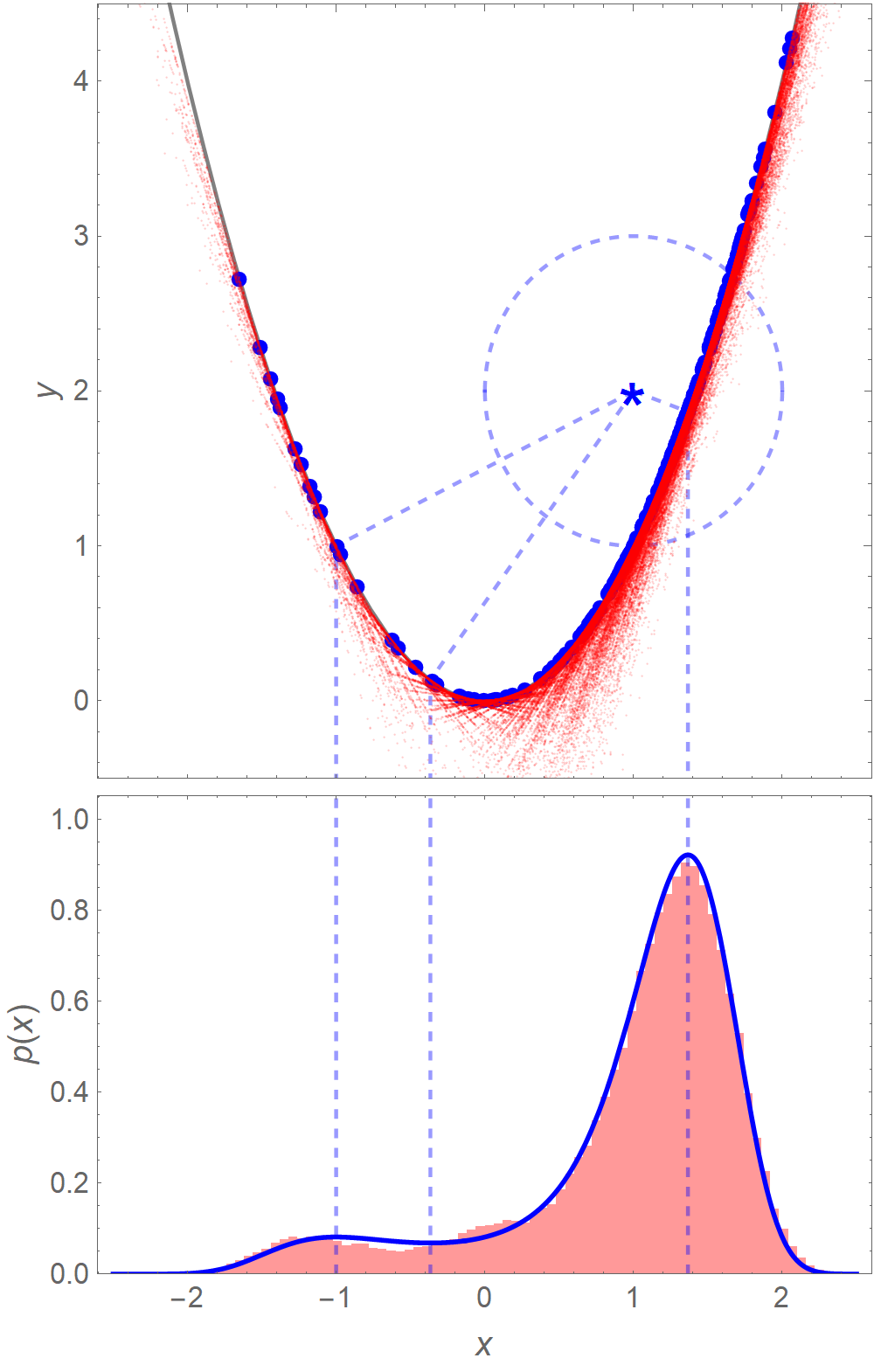}
\caption{Top: Plot of manifold $y=x^2$ (gray), base points $\alpha_i$ (blue dots), and projected points $\beta_{ij}^\perp$ (red dots). Also included are $\beta_*$ (blue star), one-sigma contour of Gaussian density $f(\beta)$ (blue circle), and stationary points of target density $p(x)$ (blue lines). Bottom: Plot of target density (blue), and histogram of projected samples $(x_{ij},w_{ij})$ (red) with bin width $0.06$.}
\label{fig:parabola}
\end{figure}

A set of $10^5$ weighted samples $(x_{ij},w_{ij})$ is generated from a base chain of length $n=200$ and mini-distributions of size $m=500$; their pushforward points $\beta_i^\perp(x_{ij})$ on the tangent bundle are displayed in the top plot of Figure \ref{fig:parabola}, together with $\beta_*$, the base points $\alpha_i$, and the manifold itself. In the bottom plot, the empirical probability density of the samples is represented by a histogram with $100$ bins, and is overlaid with the (normalized) target density \eqref{eq:parabolapdf} for comparison. The Hellinger distance between the distribution of the samples and the target distribution is $d_H=0.03$.

\begin{figure}[!tbp]
\centering
\includegraphics[width=\columnwidth]{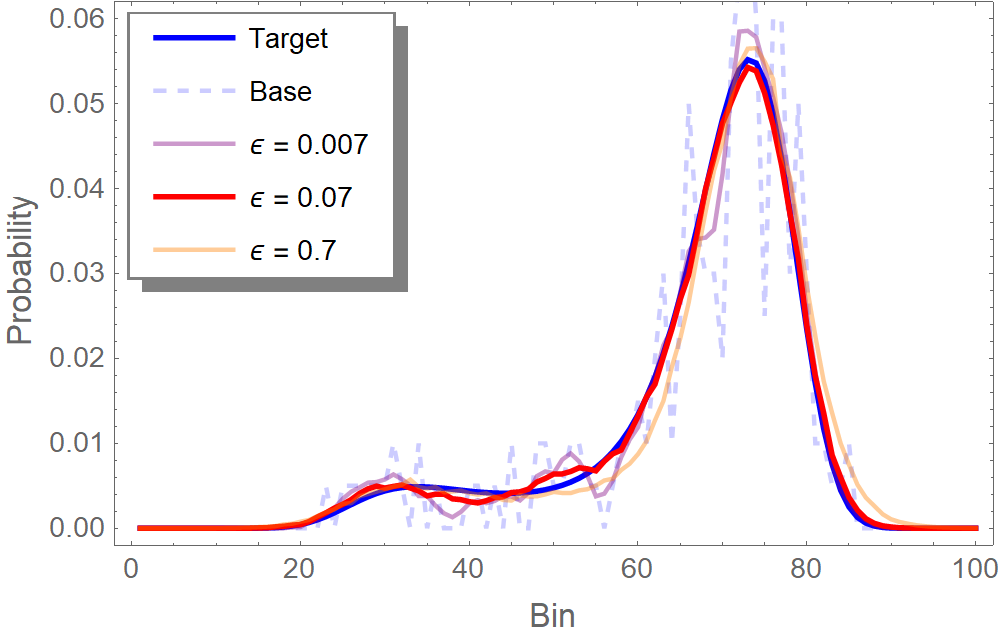}
\caption{Probability distribution of projected samples with tuning parameter $\epsilon=0.007,0.07,0.7$, along with target and base distributions. For each $\epsilon$, the Hellinger distance from the target distribution is $d_H=0.07,0.03,0.09$ respectively.}
\label{fig:epsilon}
\end{figure}

Figure \ref{fig:parabola} uses a tuning parameter of $\epsilon=0.07$, which corresponds to about $40\%$ of projected points falling within a distance of $0.07$ from the manifold (see discussion after \eqref{eq:compactness}). Although this error can be reduced with a choice of smaller $\epsilon$, the sampling resolution will become poor in the $x<0$ tail of the distribution, due to the sparsity of the base chain and the high curvature in that region. Some experimentation is needed to choose $\epsilon$ for a balance between resolution and accuracy, but finding the optimal range generally does not require fine-tuning. Figure \ref{fig:epsilon} shows the extent of changes to the histogram distribution that result from varying $\epsilon$ across two orders of magnitude. The Hellinger distance from the target distribution remains small at $d_H\leq0.09$; in comparison, the base chain has $d_H=0.22$.

We may also examine a simple test function for this example, in order to validate the estimate \eqref{eq:testerror} of the systematic error introduced by the approximate weights in the proposed method. The expectation of $x$ has the true value $\mathbf{E}[x]=0.98$; using \eqref{eq:errorweights} to compute \eqref{eq:testerror}, we obtain (for $\epsilon=0.07$) an empirical value $\mathbf{E}_q[xw]=1.01_{-0.16}$, which is consistent with the true expectation within its error. As \eqref{eq:errorweights} is constructed from a series expansion about each base sample, its accuracy should be positively correlated with the overall compactness. This is seen in the sample sets with $\epsilon=0.007,0.7$, which return $\mathbf{E}_q[xw]=0.99_{-0.01},1.05_{-0.28}$ respectively. The results for all three sets of projected samples compare reasonably to $\mathbf{E}_p[x]=1.00^{+0.06}_{-0.06}$ for the base chain, which has no systematic error (since it is generated according to $p$) and a Monte Carlo error of $\sqrt{\mathrm{Var}(x)/n}$.

\subsection{Klein bottles}
\label{subsec:klein}

Let us now consider a slightly more involved example. Our starting point is the classical embedding of a Klein bottle into $\mathbb{R}^4$ \citep{do1992riemannian}; we may promote this to a family of variable-size Klein bottles in $\mathbb{R}^5$ by setting
\begin{align}\label{eq:kleinalpha}
\alpha^1(r,\psi,\phi)&=\frac{1}{2}(1+r\cos{\psi})\cos{\phi},\nonumber\\
\alpha^2(r,\psi,\phi)&=\frac{1}{2}(1+r\cos{\psi})\sin{\phi},\nonumber\\
\alpha^3(r,\psi,\phi)&=\frac{r}{2}\sin{\psi}\cos{\frac{\phi}{2}},\nonumber\\
\alpha^4(r,\psi,\phi)&=\frac{r}{2}\sin{\psi}\sin{\frac{\phi}{2}},\nonumber\\
\alpha^5(r,\psi,\phi)&=0,
\end{align}
and then by applying an arbitrary rotation $\alpha\to R\alpha$ to bring it out of the $x^5=0$ subspace. The sampling region $\mathrm{\Theta}_b$ is taken to be the cuboid $(r,\psi,\phi)\in[2,8]\times[0,2\pi]\times[0,2\pi]$, which is an orientable three-manifold with boundary.\footnote{Admittedly, this is then not a Klein-like manifold at all---but here we are really more interested in the curvature induced by the embedding \eqref{eq:kleinalpha}.} Finally, we choose
\begin{equation}\label{eq:kleinbeta}
\beta_*=R\alpha\left(3,\frac{\pi}{4},\frac{\pi}{2}\right).
\end{equation}

The manifold $R\alpha[\mathrm{\Theta}_b]$ has non-elliptic geometry throughout, and its matrix $K$ from \eqref{eq:curvature} has rank two. Unlike the example in Section \ref{subsec:parabola}, the maximal curvature eigenvalue here is not always greater than the maximal eigenvalue of $\mathrm{\Lambda}$ from \eqref{eq:boxpushforward}, and so $c_i$ is given by \eqref{eq:simplecompactness} as designed. Three starting sets of samples are generated in this example: a reference MH chain of length $10^5$, a short base chain $B_1$ of length $n=200$, and a long base chain $B_2$ of length $n=2000$ (with $B_{1,2}$ obtained by downsampling the reference chain). Mini-distributions of size $m=500,50$ are used for $B_{1,2}$ respectively, such that the corresponding projected sets $S_{1,2}$ also have $10^5$ samples each. The tuning parameter is empirically chosen as $\epsilon=0.15,0.1$ for $S_{1,2}$ respectively. Since the target density is non-negligible at the boundary of the sampling region, any projected points $\theta_{ij}$ that fall outside $\mathrm{\Theta}_b$ are replaced with their unit-weight base samples $(\theta_i,1)$, as suggested in Section \ref{subsec:summary}.

\begin{figure}[!tbp]
\centering
\includegraphics[width=\columnwidth]{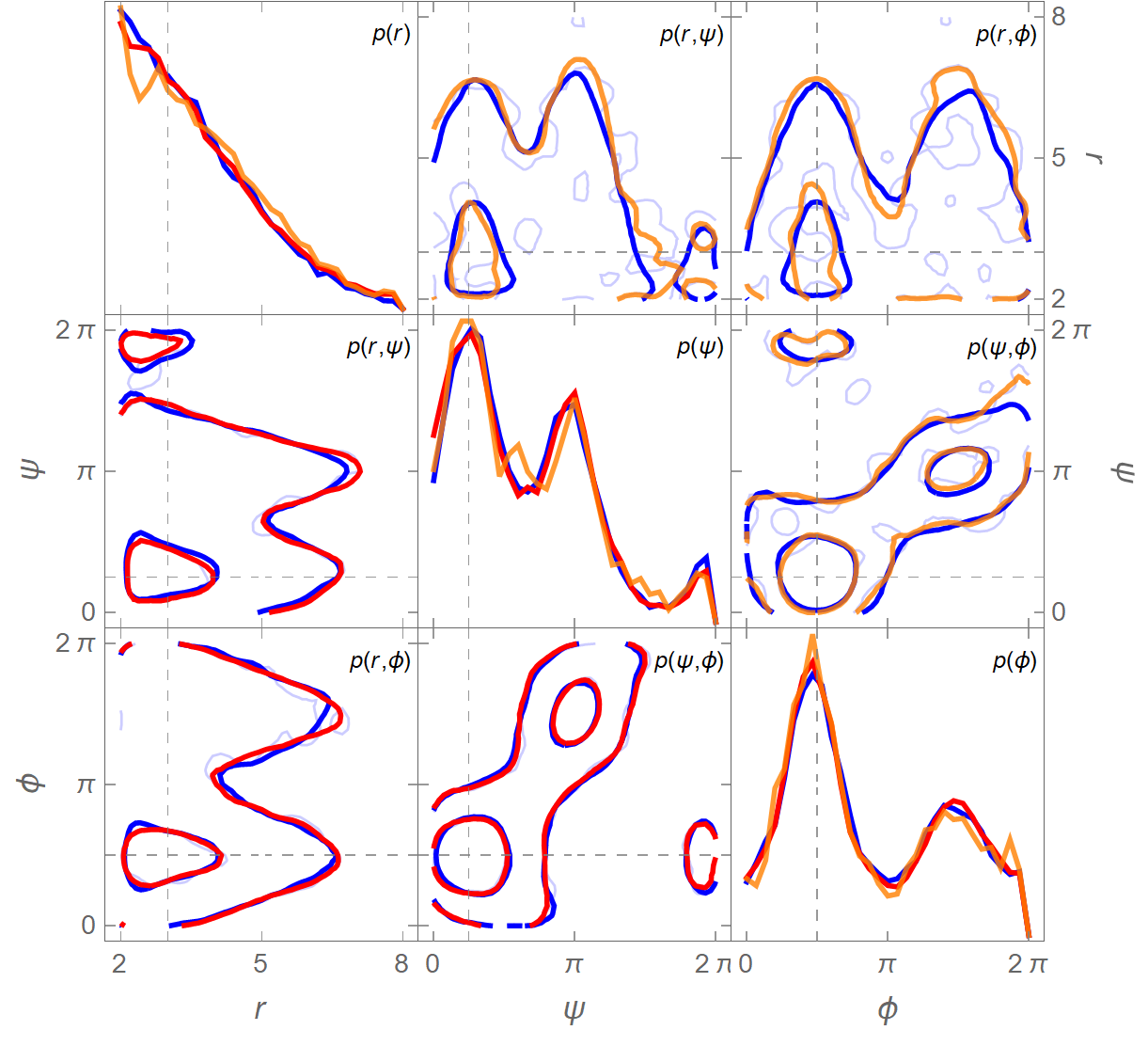}
\caption{Diagonal: Marginalized probability densities for $S_1$ (orange), $S_2$ (red), and reference chain (blue). Above diagonal: Marginalized probability density contours for $S_1$ and reference chain. Below diagonal: Marginalized probability density contours for $S_2$ and reference chain. Base chain contours (light blue) are included in contour plots. Dashed gray lines indicate location of $\beta_*$. All histograms have $30$ bins per parameter. To improve visibility, the same Gaussian smoothing kernel is applied to each contour plot. Contours are arbitrarily chosen as $p=0.02,0.08$ for illustrative purposes.}
\label{fig:klein}
\end{figure}

Probability density contours and curves (i.e., $p(r,\psi,\phi)$ marginalized over some combination of one or two parameters) are computed for $S_{1,2}$ and the reference chain, and are consolidated in Figure \ref{fig:klein}. Unsurprisingly, the Hellinger distance from the reference chain is larger for $S_1$ than it is for $S_2$; with $30$ bins per parameter (as used in the figure), it is $d_H=0.41$ compared to $d_H=0.33$. The improved accuracy of $S_2$ is also quite clear from visual inspection of the plots. Figure \ref{fig:klein} includes contours for the base chains $B_{1,2}$, which are severely under-resolved and require smoothing before they can even be visualized. The Hellinger distances for $B_{1,2}$ are $d_H=0.90,0.67$ respectively. Although the distances in this example are significantly higher than those obtained in Section \ref{subsec:parabola}, they have simply been inflated by the large number of bins. If $20$ bins per parameter are used (which is still $80$ times the number of bins in Section \ref{subsec:parabola}), the values fall to $d_H=0.30,0.21$ for $S_{1,2}$ and $d_H=0.83,0.50$ for $B_{1,2}$ respectively.

\subsection{Gravitational-wave signal space}
\label{subsec:gw}

For an example of a ``real-world'' application where the map $\alpha$ does not admit a closed-form expression (or is analytically nontrivial), we turn to the burgeoning field of gravitational-wave astronomy \citep{PhysRevLett.116.061102,PhysRevLett.119.161101,Abbott2018}. A Bayesian framework is used to conduct astrophysical inference on source signals buried in noisy detector data; in this setting, the likelihood function takes the standard form \citep{PhysRevD.49.2658,PhysRevLett.116.241102}
\begin{equation}\label{eq:gwlikelihood}
p(\theta)\propto\exp{\left(-\frac{1}{2}\left\langle h(\theta)-d_*|h(\theta)-d_*\right\rangle\right)},
\end{equation}
where $h(\theta)$ is a parametrized waveform model for the signal, $d_*$ is data comprising some ``true'' signal $h(\theta_*)$ and detector noise $n_*$, and $\langle\cdot|\cdot\rangle$ is a noise-weighted inner product that incorporates a power spectral density model for $n_*$ (assumed to be Gaussian and additive).

Eq. \eqref{eq:gwlikelihood} is manifestly a Gaussian probability density restricted to the waveform manifold $h[\mathrm{\Theta}]$, and may be cast in whitened form such that $\langle\cdot|\cdot\rangle$ simplifies to the Euclidean inner product on $\mathbb{R}^d$. However, the computation of waveform derivatives must typically be done numerically, and is compounded by the fact that $d\gtrsim10^4$ for most signals in a time- or frequency-series representation. Both of these difficulties are alleviated in recent work that uses deep-learning techniques to construct fully analytic interpolants for waveforms in a reduced-basis representation \citep{PhysRevLett.106.221102,PhysRevLett.122.211101}. Fast and reliable waveform derivatives are a key feature of this approach, in which \eqref{eq:gwlikelihood} simplifies to
\begin{equation}\label{eq:romanlikelihood}
p(\theta)\propto\exp{\left(-\frac{1}{2}\left|\alpha(\theta)-\beta_*\right|^2\right)},
\end{equation}
where $(\alpha(\theta),\beta_*)$ are obtained by projecting $(h(\theta),d_*)$ onto a compact ($d\sim10^2$) basis for $h[\mathrm{\Theta}]$.

We now apply the proposed method to the target density \eqref{eq:romanlikelihood}, where $\alpha$ is a reduced-representation interpolant for a two-parameter waveform model \citep{PhysRevD.79.104023}. This simple model describes the signal from the late inspiral stage of a non-spinning black-hole binary, with component masses in the interval $[3,30]M_\odot$ (where $M_\odot$ denotes one Solar mass); more realistic waveform models generally have $s\sim10$. The ambient dimensionality for this example is $d=173$, while the sampling region $\mathrm{\Theta}_b$ is specified in the parametrization of chirp mass $M_c$ and symmetric mass ratio $\eta$ as $(M_c,\eta)\in[7.3,8.9]M_\odot\times[0.16,0.2]$. We assume that the projected data $\beta_*$ contains a weak signal (to accentuate the features in \eqref{eq:romanlikelihood}), that the signal parameters lie at the centroid of $\mathrm{\Theta}_b$, and that the projection of the detector noise onto the reduced basis vanishes, i.e.,
\begin{equation}\label{eq:gwbeta}
\beta_*=\alpha(8.1M_\odot,0.18).
\end{equation}

\begin{figure}[!tbp]
\centering
\includegraphics[width=1.1\columnwidth]{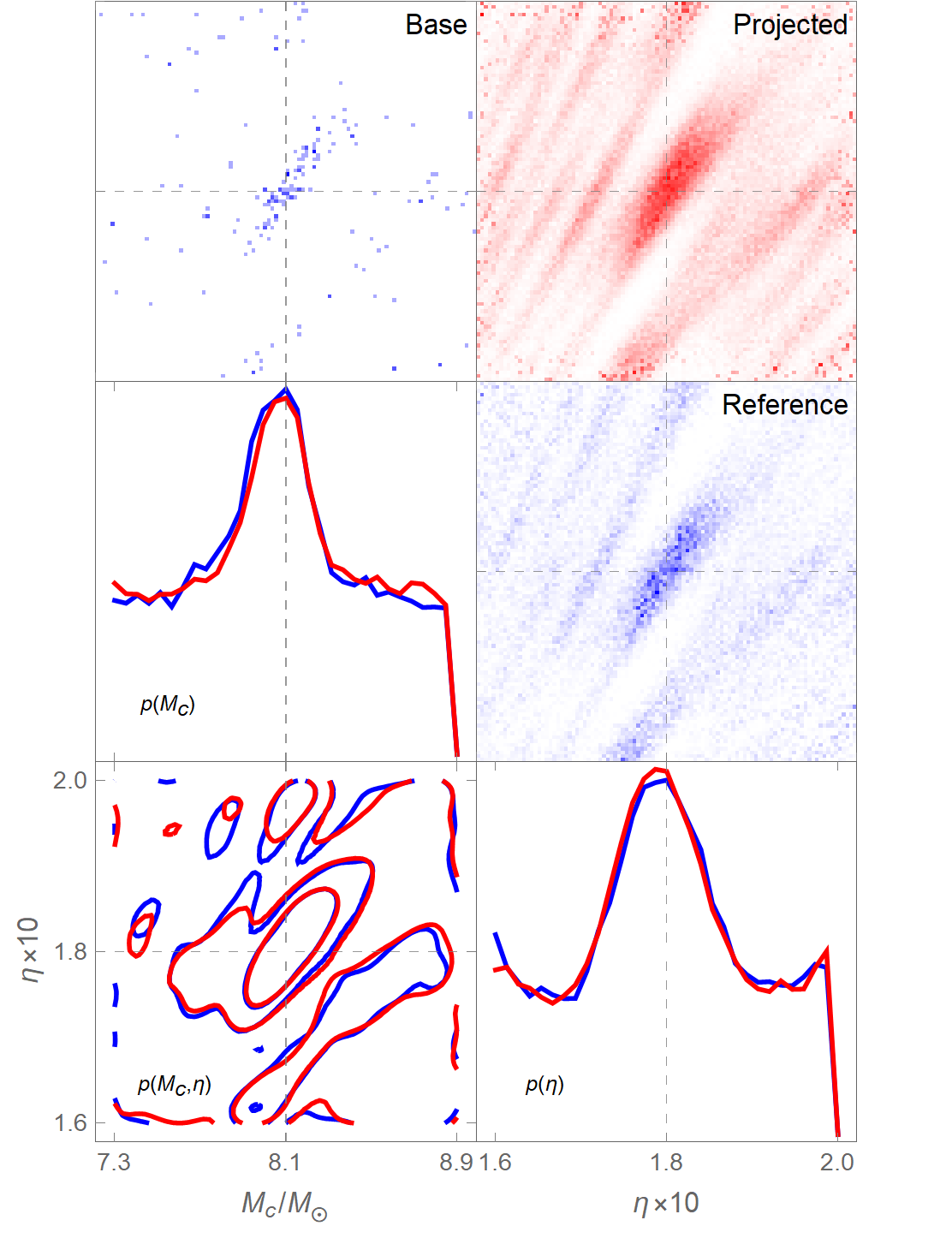}
\caption{Top and right: High-resolution density histograms for base chain, projected samples, and reference chain. Dashed gray lines indicate location of $\beta_*$. Histograms have $100$ bins per parameter. Bottom and left: Marginalized probability densities and density contours for projected samples (red) and reference chain (blue). Histograms have $30$ bins per parameter. To improve visibility, the same Gaussian smoothing kernel is applied to each contour plot. Contours are arbitrarily chosen as $p=0.5,1.8,4.0$ for illustrative purposes.}
\label{fig:gw}
\end{figure}

The MH algorithm is used to draw $1.5\times10^5$ samples according to the density \eqref{eq:romanlikelihood}; this reference chain is then downsampled to a base chain of length $n=1500$. Both $\lambda_i^2$ and $\kappa_i$ in \eqref{eq:simplecompactness} turn out to vary by less than a factor of five over all points in the base chain. It is instructive to examine the robustness of the proposed method when the compactness is approximated as constant in such cases, even though the second derivative of $\alpha$ is available. Setting $m=100$, $c_i=1/\epsilon$ and $\epsilon=0.7$, we obtain a projected set of $1.5\times10^5$ weighted samples in $\mathrm{\Theta}_b$ (after applying boundary corrections as in Section \ref{subsec:klein}). With $30$ bins per parameter, the Hellinger distance from the reference chain is $d_H=0.10$ for the projected samples, and $d_H=0.33$ for the base chain. The visual agreement between the projected samples and the reference chain (especially in the high-resolution histograms of Figure \ref{fig:gw}) is also serviceable for most intents and purposes, notwithstanding the constant compactness and the increased complexity of the probability surface.

\subsection{Parabola redux}
\label{subsec:reparabola}

It is tempting to investigate, at least for simple problems, whether a map $\alpha':\mathrm{\Theta}\to\mathbb{R}^d$ can be constructed such that (i) the image manifold $\alpha'[\mathrm{\Theta}]$ has vanishing normal curvature, but (ii) the target density $p(\theta)\propto f(\alpha'(\theta))$ is preserved with respect to the Lebesgue measure on $\mathrm{\Theta}$. As a cost of fulfilling these two conditions, we will give up the requirement that $\alpha'$ is an embedding or even an immersion, i.e., the pushforward of $\alpha'$ is no longer assumed to be injective for all $\theta\in\mathrm{\Theta}$. Then the sample-space manifold $(\mathrm{\Theta},\mathrm{I})$ is not generally Riemannian, as the pullback metric might not be positive-definite and can vanish at certain singular points on $\mathrm{\Theta}$. From a practical standpoint, however, a reparametrization of the problem with vanishing $\kappa_i$ allows the compactness to be computed from the metric term $\lambda_i^2$ alone, and the existence of singular points is not necessarily an issue if the set of all such points has Lebesgue measure zero.

Let us return to the parabola-restricted Gaussian in Section \ref{subsec:parabola}, and find some alternative $(\alpha(\xi),\beta_*)$ such that $\kappa_i=0$ in \eqref{eq:simplecompactness} and $p(\xi)$ has the same functional form as \eqref{eq:parabolapdf}. One possible choice is
\begin{equation}\label{eq:reparabolaalpha}
\alpha(\xi)=\left(\sqrt{\xi^4-3\xi^2-2\xi+5},0\right),
\end{equation}
\begin{equation}\label{eq:reparabolabeta}
\beta_*=(0,0).
\end{equation}
The image of the map \eqref{eq:reparabolaalpha} is then a ray on the $x^1$-axis:
\begin{equation}\label{eq:ray}
\alpha^1(\xi)\in\left[\frac{1}{2}\sqrt{11-6\sqrt{3}},\infty\right),
\end{equation}
while the single metric component
\begin{equation}\label{eq:reparabolametric}
F_\mathrm{I}(\xi)=J^1_{\;1}(\xi)^2=\frac{\left(2\xi^3-3\xi-1\right)^2}{\alpha^1(\xi)^2}
\end{equation}
vanishes at $\xi_s=-1,(1\pm\sqrt{3})/2$. With $\xi_i\in[-3,3]$ as before, the metric term in \eqref{eq:simplecompactness} is given by
\begin{equation}\label{eq:reparabolacompactness}
\lambda_i^2=\frac{1}{9F_\mathrm{I}(\xi_i)}.
\end{equation}

\begin{figure}[!tbp]
\centering
\includegraphics[width=\columnwidth]{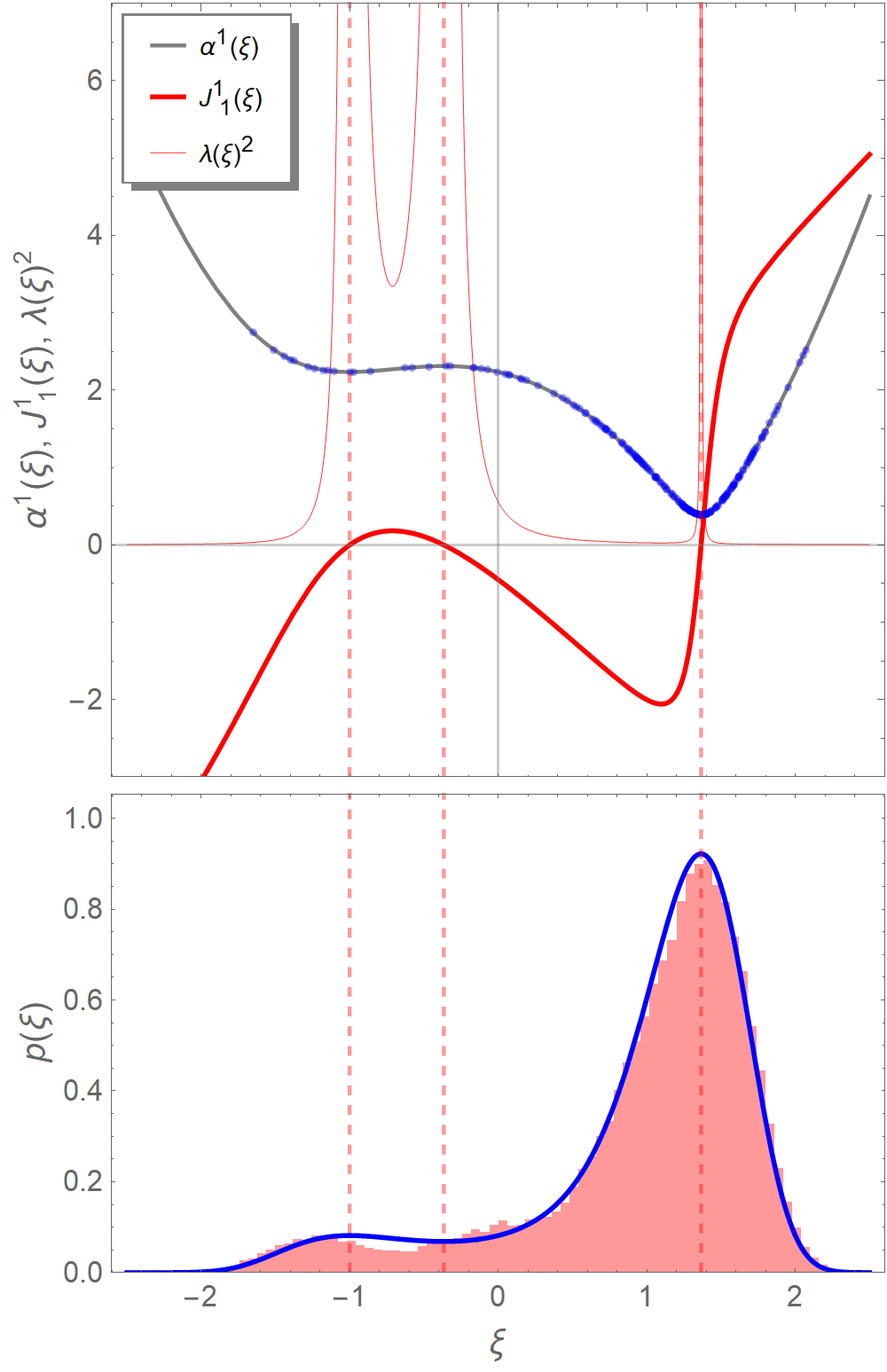}
\caption{Top: Plot of (nonzero components of) $\alpha(\xi)$ and $J(\xi)$, and compactness term $\lambda(\xi)^2$. Also included are base points $\alpha_i$ (blue dots) and singular points of metric $F_\mathrm{I}(\xi)$ (red lines). Bottom: Plot of target density $p(\xi)$ (blue), and histogram of projected samples $(\xi_{ij},w_{ij})$ (red) with bin width $0.06$.}
\label{fig:reparabola}
\end{figure}

As seen from the top plot of Figure \ref{fig:reparabola}, the singular points $\xi_s$ are precisely the stationary points of $p(\xi)$, and the compactness diverges at the corresponding points $\alpha_s$, such that $\beta_{ij}^\perp\to\alpha_i$ for base points near $\alpha_s$. This high compactness compensates for the wide dispersion of pullback samples due to the vanishing metric. Note that for the singular point corresponding to the global minimum of $\alpha^1$, the $x^1$-coordinate of a projected point $\beta_{ij}^\perp$ near $\alpha_s$ can fall outside the interval in \eqref{eq:ray}, but both the pullback sample $\xi_{ij}$ and its weight $w_{ij}$ remain well defined in such cases. The metric-only compactness is also less effective near this singular point since the metric changes rapidly there, and hence a deficit of samples in that region might occur for a sparse base chain. Regardless, the bottom plot of Figure \ref{fig:reparabola} shows that the proposed method still works with $(n,m)=(200,500)$ (i.e., the same base chain as in Section \ref{subsec:parabola}), but the tuning parameter must be reduced to $\epsilon=0.005$ because of the reparametrization. Using a bin width of $0.06$, the Hellinger distance between the target distribution and that of the projected samples is $d_H=0.04$.

\subsection{Beta distribution}
\label{subsec:beta}

As a final example, let us consider a broader class of target distributions that are not manifestly restricted Gaussians, but might be cast as such through the reparametrization approach taken in Section \ref{subsec:reparabola}. We begin, rather ambitiously, with a probability distribution from a general exponential family \citep{brown1986fundamentals}. This covers any distribution whose density function can be written as
\begin{equation}\label{eq:exponentialfamily}
p(\theta)=\exp{\left(T(\theta)+V(\theta)^TW(\zeta)+Z(\zeta)\right)},
\end{equation}
where $T,Z$ are scalar-valued, $V,W$ are vector-valued with some dimensionality $k$, and the vector $\zeta$ parametrizes the particular exponential family defined by the functional forms of $T,V,W,Z$.

For fixed $\zeta$, we may cast \eqref{eq:exponentialfamily} in the restricted-Gaussian form $p(\theta)\propto f(\alpha(\theta))$, through a crude but direct choice of $\alpha$ (with $\beta_*=0$):
\begin{align}\label{eq:exponentialalphabeta}
\alpha^1(\theta)&=\sqrt{-2T(\theta)},\nonumber\\
\alpha^2(\theta)&=\sqrt{-2W^1V^1(\theta)},\nonumber\\
&\;\;\vdots\nonumber\\
\alpha^d(\theta)&=\sqrt{-2W^kV^k(\theta)},
\end{align}
where $d=k+1$.\footnote{By definition, we have $k=s$ for a full exponential family and $k>s$ for a curved one, such that $d>s$ as required.} It is clear that such a map will in general be complex-valued (which then requires lifting the whole framework to complex space), and might also give rise to an induced metric with pathologies (e.g., vanishing or divergent points, non-smoothness, etc.)

Many common probability distributions are exponential families; continuous examples include the normal, gamma and chi-squared distributions. Our focus in this section will be restricted to the beta distribution, which is defined on the interval $[0,1]$ and has two positive shape parameters $(a,b)$. Following \eqref{eq:exponentialalphabeta}, we set
\begin{align}\label{eq:betaalpha}
\alpha^1(\xi)&=\sqrt{2\ln{(\xi(1-\xi))}},\nonumber\\
\alpha^2(\xi)&=\sqrt{-2a\ln{\xi}},\nonumber\\
\alpha^3(\xi)&=\sqrt{-2b\ln{(1-\xi)}},
\end{align}
such that the target density is (as desired)
\begin{equation}\label{eq:betapdf}
p(\xi)\propto\xi^{a-1}(1-\xi)^{b-1}.
\end{equation}

Observe that $\alpha^1$ is imaginary and $\alpha^{2,3}$ are real for all $\xi\in(0,1)$, with singular points only at the boundary. The image of $\alpha$ is then confined to a subspace of $\mathbb{C}^3$ that is diffeomorphic to a subspace of $\mathbb{R}^3$, since the linear transformation defined by the matrix $\mathrm{diag}(i,1,1)$ is a smooth bijection between said subspaces. We are then justified in continuing to treat the ambient space as $\mathbb{R}^3$ instead of $\mathbb{C}^3$, by not promoting the ambient normal densities associated with \eqref{eq:gaussiansamples} and \eqref{eq:gaussiandensity} to their complex counterparts. This approach allows the framework in Section \ref{sec:method} to be used largely unchanged, with one caveat: the metric is degenerate with varying signature for certain values of the shape parameters. As the manifold is still locally pseudo-Riemannian, we get around this by simply taking the absolute value of any negative metric and curvature terms in \eqref{eq:simplecompactness}. Explicit use of the above bijection is only required when standard computational routines are limited to real inputs and outputs, e.g., the generation of random variates.

\begin{figure}[!tbp]
\centering
\includegraphics[width=\columnwidth]{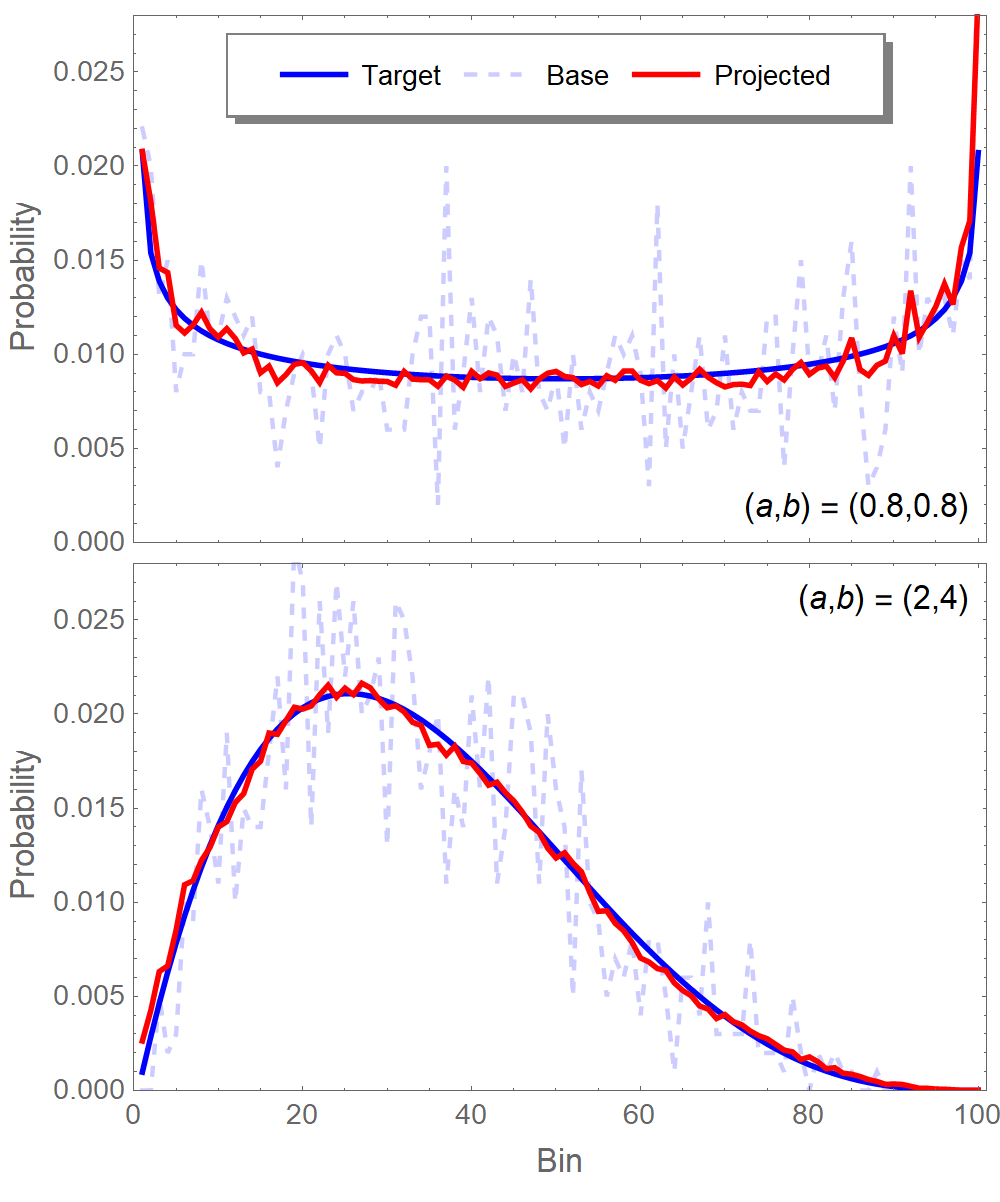}
\caption{Top: Probability distribution of projected samples for $(a,b)=(0.8,0.8)$, along with target and base distributions. Bottom: As above, but for $(a,b)=(2,4)$.}
\label{fig:beta}
\end{figure}

With these adjustments, the proposed method is employed to sample from two qualitatively different beta distributions: one with shape parameters $(a,b)=(0.8,0.8)$, and the other with $(a,b)=(2,4)$. Each set of projected samples is generated from a base chain of length $n=1000$ and mini-distributions of size $m=100$. The tuning parameter is chosen as $\epsilon=0.15$ for $(a,b)=(0.8,0.8)$, and $\epsilon=0.03$ for $(a,b)=(2,4)$. Figure \ref{fig:beta} shows the histogram distributions for both sample sets, with the corresponding target and base distributions overlaid for comparison (all histograms have a bin width of $0.01$). For $(a,b)=(0.8,0.8)$, the Hellinger distance from the target distribution is $d_H=0.12$ for the base chain, and falls to $d_H=0.03$ for the projected samples; for $(a,b)=(2,4)$, it decreases from $d_H=0.13$ to $d_H=0.03$. Although the projected sample sets appear to be slightly under-resolved (i.e., over-correlated with their base chains), $\epsilon$ cannot be raised much further in this example without incurring a significant penalty to sampling accuracy, and so a longer base chain is necessary if increased smoothing is desired.

\section{Conclusion}
\label{sec:conclusion}

In this paper, I have proposed a derivative-based framework for sampling approximately and efficiently from continuous probability distributions that admit a density of the form \eqref{eq:pdf}. These are essentially standard closed-form distributions on Euclidean space, but restricted to a parametrized submanifold of their domain. Much of the present work deals with the specific but important case where the ambient-space distribution is a multivariate Gaussian. Examples in Sections \ref{subsec:reparabola} and \ref{subsec:beta}, while somewhat academic, indicate that reparametrization might be a viable strategy for generalizing the existing results to a broader class of distributions (and their manifold restrictions, by extension).

The envisioned utility of the proposed method lies primarily in facilitating density estimation for Bayesian inference problems, and less so in MCMC integration (due to the systematic error \eqref{eq:testerror}). It multiplicatively upsamples the output of any standard algorithm that might be employed in such applications, and is particularly synergistic with manifold-based samplers. Only the first derivative of the smooth map is required for the construction of projected samples, although second-derivative information can be incorporated naturally and to good effect. The method is robust, with only a single tuning parameter for the overall spread of the mini-distributions. Its end product is a high-resolution set of weighted samples that can be used to build histograms, or more generally to enable kernel density estimation with a smaller optimal bandwidth.

\section*{Acknowledgments}

This manuscript was graced with the most incisive and constructive review I have ever received, for which I am indebted to an anonymous Statistics and Computing referee. I am also grateful for pivotal discussions with Michele Vallisneri and his scrutiny of an early draft, as well as Christopher Moore's insight and detailed comments on the method. Final thanks go out to Chad Galley, Christopher Berry, Jonathan Gair and Leo Stein for helpful conversations and suggestions. This work was supported by the JPL Research and Technology Development program, and was carried out at JPL, California Institute of Technology, under a contract with the National Aeronautics and Space Administration. \copyright\,2019 California Institute of Technology. U.S. Government sponsorship acknowledged.

\bibliographystyle{chicago}
\bibliography{references}

\begin{thebibliography}{}

\bibitem[\protect\citeauthoryear{Abbott et~al.}{Abbott
  et~al.}{2016a}]{PhysRevLett.116.061102}
Abbott, B.~P. et~al. (2016a, Feb).
\newblock Observation of gravitational waves from a binary black hole merger.
\newblock {\em Phys. Rev. Lett.\/}~{\em 116}, 061102.

\bibitem[\protect\citeauthoryear{Abbott et~al.}{Abbott
  et~al.}{2016b}]{PhysRevLett.116.241102}
Abbott, B.~P. et~al. (2016b, Jun).
\newblock Properties of the binary black hole merger gw150914.
\newblock {\em Phys. Rev. Lett.\/}~{\em 116}, 241102.

\bibitem[\protect\citeauthoryear{Abbott et~al.}{Abbott
  et~al.}{2017}]{PhysRevLett.119.161101}
Abbott, B.~P. et~al. (2017, Oct).
\newblock Gw170817: Observation of gravitational waves from a binary neutron
  star inspiral.
\newblock {\em Phys. Rev. Lett.\/}~{\em 119}, 161101.

\bibitem[\protect\citeauthoryear{Abbott et~al.}{Abbott
  et~al.}{2018}]{Abbott2018}
Abbott, B.~P. et~al. (2018, Apr).
\newblock Prospects for observing and localizing gravitational-wave transients
  with advanced ligo, advanced virgo and kagra.
\newblock {\em Living Reviews in Relativity\/}~{\em 21\/}(1), 3.

\bibitem[\protect\citeauthoryear{Amari and Nagaoka}{Amari and
  Nagaoka}{2007}]{amari2007methods}
Amari, S. and H.~Nagaoka (2007).
\newblock {\em Methods of Information Geometry}.
\newblock Translations of mathematical monographs. American Mathematical
  Society.

\bibitem[\protect\citeauthoryear{Arun, Buonanno, Faye, and Ochsner}{Arun
  et~al.}{2009}]{PhysRevD.79.104023}
Arun, K.~G., A.~Buonanno, G.~Faye, and E.~Ochsner (2009, May).
\newblock Higher-order spin effects in the amplitude and phase of gravitational
  waveforms emitted by inspiraling compact binaries: Ready-to-use gravitational
  waveforms.
\newblock {\em Phys. Rev. D\/}~{\em 79}, 104023.

\bibitem[\protect\citeauthoryear{Arvanitidis, Hansen, and Hauberg}{Arvanitidis
  et~al.}{2018}]{arvanitidis2018latent}
Arvanitidis, G., L.~K. Hansen, and S.~Hauberg (2018).
\newblock Latent space oddity: on the curvature of deep generative models.
\newblock In {\em International Conference on Learning Representations}.

\bibitem[\protect\citeauthoryear{Bishop}{Bishop}{2013}]{bishop2013pattern}
Bishop, C. (2013).
\newblock {\em Pattern Recognition and Machine Learning}.
\newblock Information science and statistics. Springer (India) Private Limited.

\bibitem[\protect\citeauthoryear{Brown, of~Mathematical~Statistics, and
  (Organization)}{Brown et~al.}{1986}]{brown1986fundamentals}
Brown, L., I.~of~Mathematical~Statistics, and J.~(Organization) (1986).
\newblock {\em Fundamentals of Statistical Exponential Families: With
  Applications in Statistical Decision Theory}.
\newblock IMS Lecture Notes. Institute of Mathematical Statistics.

\bibitem[\protect\citeauthoryear{Brubaker, Salzmann, and Urtasun}{Brubaker
  et~al.}{2012}]{pmlr-v22-brubaker12}
Brubaker, M., M.~Salzmann, and R.~Urtasun (2012, 21--23 Apr).
\newblock A family of mcmc methods on implicitly defined manifolds.
\newblock In N.~D. Lawrence and M.~Girolami (Eds.), {\em Proceedings of the
  Fifteenth International Conference on Artificial Intelligence and
  Statistics}, Volume~22 of {\em Proceedings of Machine Learning Research}, La
  Palma, Canary Islands, pp.\  161--172. PMLR.

\bibitem[\protect\citeauthoryear{Chua, Galley, and Vallisneri}{Chua
  et~al.}{2019}]{PhysRevLett.122.211101}
Chua, A. J.~K., C.~R. Galley, and M.~Vallisneri (2019, May).
\newblock Reduced-order modeling with artificial neurons for gravitational-wave
  inference.
\newblock {\em Phys. Rev. Lett.\/}~{\em 122}, 211101.

\bibitem[\protect\citeauthoryear{Cutler and Flanagan}{Cutler and
  Flanagan}{1994}]{PhysRevD.49.2658}
Cutler, C. and E.~E. Flanagan (1994, Mar).
\newblock Gravitational waves from merging compact binaries: How accurately can
  one extract the binary's parameters from the inspiral waveform?
\newblock {\em Phys. Rev. D\/}~{\em 49}, 2658--2697.

\bibitem[\protect\citeauthoryear{Diaconis, Holmes, and Shahshahani}{Diaconis
  et~al.}{2013}]{diaconis}
Diaconis, P., S.~Holmes, and M.~Shahshahani (2013).
\newblock {\em Sampling from a Manifold}, Volume Volume 10 of {\em
  Collections}, pp.\  102--125.
\newblock Beachwood, Ohio, USA: Institute of Mathematical Statistics.

\bibitem[\protect\citeauthoryear{do~Carmo}{do~Carmo}{1992}]{do1992riemannian}
do~Carmo, M. (1992).
\newblock {\em Riemannian Geometry}.
\newblock Mathematics (Boston, Mass.). Birkh{\"a}user.

\bibitem[\protect\citeauthoryear{Duane, Kennedy, Pendleton, and Roweth}{Duane
  et~al.}{1987}]{DUANE1987216}
Duane, S., A.~D. Kennedy, B.~J. Pendleton, and D.~Roweth (1987).
\newblock Hybrid monte carlo.
\newblock {\em Physics Letters B\/}~{\em 195\/}(2), 216 -- 222.

\bibitem[\protect\citeauthoryear{Field, Galley, Herrmann, Hesthaven, Ochsner,
  and Tiglio}{Field et~al.}{2011}]{PhysRevLett.106.221102}
Field, S.~E., C.~R. Galley, F.~Herrmann, J.~S. Hesthaven, E.~Ochsner, and
  M.~Tiglio (2011, Jun).
\newblock Reduced basis catalogs for gravitational wave templates.
\newblock {\em Phys. Rev. Lett.\/}~{\em 106}, 221102.

\bibitem[\protect\citeauthoryear{Gamerman and Lopes}{Gamerman and
  Lopes}{2006}]{gamerman2006markov}
Gamerman, D. and H.~Lopes (2006).
\newblock {\em Markov Chain Monte Carlo: Stochastic Simulation for Bayesian
  Inference, Second Edition}.
\newblock Chapman \& Hall/CRC Texts in Statistical Science. Taylor \& Francis.

\bibitem[\protect\citeauthoryear{Gelman, Carlin, Stern, Dunson, Vehtari, and
  Rubin}{Gelman et~al.}{2013}]{gelman2013bayesian}
Gelman, A., J.~Carlin, H.~Stern, D.~Dunson, A.~Vehtari, and D.~Rubin (2013).
\newblock {\em Bayesian Data Analysis, Third Edition}.
\newblock Chapman \& Hall/CRC Texts in Statistical Science. CRC Press.

\bibitem[\protect\citeauthoryear{Girolami and Calderhead}{Girolami and
  Calderhead}{2011}]{doi:10.1111/j.1467-9868.2010.00765.x}
Girolami, M. and B.~Calderhead (2011).
\newblock Riemann manifold langevin and hamiltonian monte carlo methods.
\newblock {\em Journal of the Royal Statistical Society: Series B (Statistical
  Methodology)\/}~{\em 73\/}(2), 123--214.

\bibitem[\protect\citeauthoryear{Golub and Kahan}{Golub and
  Kahan}{1965}]{golub1965calculating}
Golub, G. and W.~Kahan (1965).
\newblock Calculating the singular values and pseudo-inverse of a matrix.
\newblock {\em Journal of the Society for Industrial and Applied Mathematics,
  Series B: Numerical Analysis\/}~{\em 2\/}(2), 205--224.

\bibitem[\protect\citeauthoryear{Graff, Hobson, and Lasenby}{Graff
  et~al.}{2011}]{10.1111/j.1745-3933.2011.01034.x}
Graff, P., M.~P. Hobson, and A.~Lasenby (2011, 05).
\newblock {An investigation into the Multiple Optimised Parameter Estimation
  and Data compression algorithm}.
\newblock {\em Monthly Notices of the Royal Astronomical Society:
  Letters\/}~{\em 413\/}(1), L66--L70.

\bibitem[\protect\citeauthoryear{Hastings}{Hastings}{1970}]{doi:10.1093/biomet/57.1.97}
Hastings, W.~K. (1970).
\newblock Monte carlo sampling methods using markov chains and their
  applications.
\newblock {\em Biometrika\/}~{\em 57\/}(1), 97--109.

\bibitem[\protect\citeauthoryear{Heavens, Jimenez, and Lahav}{Heavens
  et~al.}{2000}]{10.1046/j.1365-8711.2000.03692.x}
Heavens, A.~F., R.~Jimenez, and O.~Lahav (2000, 10).
\newblock {Massive lossless data compression and multiple parameter estimation
  from galaxy spectra}.
\newblock {\em Monthly Notices of the Royal Astronomical Society\/}~{\em
  317\/}(4), 965--972.

\bibitem[\protect\citeauthoryear{Hjort, Holmes, M{\"u}ller, and Walker}{Hjort
  et~al.}{2010}]{hjort2010bayesian}
Hjort, N., C.~Holmes, P.~M{\"u}ller, and S.~Walker (2010).
\newblock {\em Bayesian Nonparametrics}.
\newblock Cambridge Series in Statistical and Probabilistic Mathematics.
  Cambridge University Press.

\bibitem[\protect\citeauthoryear{Kessy, Lewin, and Strimmer}{Kessy
  et~al.}{2018}]{doi:10.1080/00031305.2016.1277159}
Kessy, A., A.~Lewin, and K.~Strimmer (2018).
\newblock Optimal whitening and decorrelation.
\newblock {\em The American Statistician\/}~{\em 0\/}(0), 1--6.

\bibitem[\protect\citeauthoryear{Kotz, Balakrishnan, and Johnson}{Kotz
  et~al.}{2004}]{kotz2004continuous}
Kotz, S., N.~Balakrishnan, and N.~Johnson (2004).
\newblock {\em Continuous Multivariate Distributions, Volume 1: Models and
  Applications}.
\newblock Continuous Multivariate Distributions. Wiley.

\bibitem[\protect\citeauthoryear{Lee and Verleysen}{Lee and
  Verleysen}{2007}]{lee2007nonlinear}
Lee, J. and M.~Verleysen (2007).
\newblock {\em Nonlinear Dimensionality Reduction}.
\newblock Information Science and Statistics. Springer New York.

\bibitem[\protect\citeauthoryear{{Leli{\`e}vre}, {Rousset}, and
  {Stoltz}}{{Leli{\`e}vre} et~al.}{2018}]{2018arXiv180702356L}
{Leli{\`e}vre}, T., M.~{Rousset}, and G.~{Stoltz} (2018, Jul).
\newblock {Hybrid Monte Carlo methods for sampling probability measures on
  submanifolds}.
\newblock {\em arXiv e-prints\/}, arXiv:1807.02356.

\bibitem[\protect\citeauthoryear{Ma, Chen, and Fox}{Ma
  et~al.}{2015}]{Ma:2015:CRS:2969442.2969566}
Ma, Y.-A., T.~Chen, and E.~B. Fox (2015).
\newblock A complete recipe for stochastic gradient mcmc.
\newblock In {\em Proceedings of the 28th International Conference on Neural
  Information Processing Systems - Volume 2}, NIPS'15, Cambridge, MA, USA, pp.\
   2917--2925. MIT Press.

\bibitem[\protect\citeauthoryear{Meyer}{Meyer}{2000}]{meyer2000matrix}
Meyer, C. (2000).
\newblock {\em Matrix Analysis and Applied Linear Algebra:}.
\newblock Other Titles in Applied Mathematics. Society for Industrial and
  Applied Mathematics.

\bibitem[\protect\citeauthoryear{Peskin and Schroeder}{Peskin and
  Schroeder}{1995}]{peskin1995introduction}
Peskin, M. and D.~Schroeder (1995).
\newblock {\em An Introduction To Quantum Field Theory}.
\newblock Frontiers in Physics. Avalon Publishing.

\bibitem[\protect\citeauthoryear{Pollard}{Pollard}{2002}]{pollard2002user}
Pollard, D. (2002).
\newblock {\em A User's Guide to Measure Theoretic Probability}.
\newblock Cambridge Series in Statistical and Probabilistic Mathematics.
  Cambridge University Press.

\bibitem[\protect\citeauthoryear{Press, Teukolsky, Vetterling, and
  Flannery}{Press et~al.}{2007}]{press2007numerical}
Press, W., S.~Teukolsky, W.~Vetterling, and B.~Flannery (2007).
\newblock {\em Numerical Recipes 3rd Edition: The Art of Scientific Computing}.
\newblock Cambridge University Press.

\bibitem[\protect\citeauthoryear{Protopapas, Jimenez, and Alcock}{Protopapas
  et~al.}{2005}]{10.1111/j.1365-2966.2005.09305.x}
Protopapas, P., R.~Jimenez, and C.~Alcock (2005, 09).
\newblock {Fast identification of transits from light-curves}.
\newblock {\em Monthly Notices of the Royal Astronomical Society\/}~{\em
  362\/}(2), 460--468.

\bibitem[\protect\citeauthoryear{Ripley}{Ripley}{1987}]{ripley1987stochastic}
Ripley, B. (1987).
\newblock {\em Stochastic simulation}.
\newblock Wiley Series in Probability and Statistics. J. Wiley.

\bibitem[\protect\citeauthoryear{Robert and Casella}{Robert and
  Casella}{2005}]{robert2005monte}
Robert, C. and G.~Casella (2005).
\newblock {\em Monte Carlo Statistical Methods}.
\newblock Springer Texts in Statistics. Springer New York.

\bibitem[\protect\citeauthoryear{Roberts and Rosenthal}{Roberts and
  Rosenthal}{1998}]{doi:10.1111/1467-9868.00123}
Roberts, G.~O. and J.~S. Rosenthal (1998).
\newblock Optimal scaling of discrete approximations to langevin diffusions.
\newblock {\em Journal of the Royal Statistical Society: Series B (Statistical
  Methodology)\/}~{\em 60\/}(1), 255--268.

\bibitem[\protect\citeauthoryear{Rosenblatt}{Rosenblatt}{1952}]{rosenblatt1952}
Rosenblatt, M. (1952, 09).
\newblock Remarks on a multivariate transformation.
\newblock {\em Ann. Math. Statist.\/}~{\em 23\/}(3), 470--472.

\bibitem[\protect\citeauthoryear{Shephard}{Shephard}{1991}]{shephard_1991}
Shephard, N. (1991).
\newblock From characteristic function to distribution function: A simple
  framework for the theory.
\newblock {\em Econometric Theory\/}~{\em 7\/}(4), 519–529.

\bibitem[\protect\citeauthoryear{Spivak}{Spivak}{1970}]{spivak1970comprehensive}
Spivak, M. (1970).
\newblock {\em A comprehensive introduction to differential geometry}.
\newblock Number v. 1 in A Comprehensive Introduction to Differential Geometry.
  Brandeis University.

\bibitem[\protect\citeauthoryear{Struik}{Struik}{1961}]{struik1961lectures}
Struik, D. (1961).
\newblock {\em Lectures on Classical Differential Geometry}.
\newblock Addison-Wesley series in mathematics. Addison-Wesley Publishing
  Company.

\bibitem[\protect\citeauthoryear{Xifara, Sherlock, Livingstone, Byrne, and
  Girolami}{Xifara et~al.}{2014}]{XIFARA201414}
Xifara, T., C.~Sherlock, S.~Livingstone, S.~Byrne, and M.~Girolami (2014).
\newblock Langevin diffusions and the metropolis-adjusted langevin algorithm.
\newblock {\em Statistics \& Probability Letters\/}~{\em 91}, 14 -- 19.

\end{thebibliography}

\end{document}